\documentclass[aps,twocolumn,prb,showpacs,tightenlines]{revtex4}
\usepackage{amssymb}
\usepackage{amsmath}
\usepackage{graphicx}
\usepackage{epsfig}
\usepackage{txfonts}
\usepackage{color}

\begin{document}

\title{Quantum noise theory for quantum transport through nanostructures}
\author{Nan Zhao}
\author{Jia-Lin Zhu}
\affiliation{Department of Physics, Tsinghua University, Beijing 100084, China}
\author{R.-B. Liu}
\affiliation{Department of Physics, The Chinese University of Hong Kong, Shatin, N. T.,
Hong Kong, China}
\author{C. P. Sun}
\affiliation{Institute of Theoretical Physics, Chinese Academy of Sciences, Beijing,
100080, China}

\begin{abstract}
We develop a quantum noise approach to study quantum transport
through nanostructures. The nanostructures, such as quantum dots,
are regarded as artificial atoms, subject to quasi-equilibrium
fermionic reservoirs of electrons in biased leads. Noise operators
characterizing the quantum fluctuation in the reservoirs are related
to the damping and fluctuation of the artificial atoms through the
quantum Langevin equation. The average current and current noise are
derived in terms of the reservoir noise correlations. In the
white-noise limit, we show that the current and current noise can be
exactly calculated by the quantum noise approach, even in the
presence of interaction such as Coulomb blockade. As a typical application, 
the average current and current noise through a single quantum dot
are studied.
\end{abstract}

\pacs{
73.63.Kv, 
73.23.Hk, 42.50.Lc
}
\maketitle

\section{Introduction}

Quantum transport through nanostructures is of importance in
nano-science and nano-technology. Many electronic devices based on
nanostructures, such as single-electron transistors, have been
studied in the past decades for their potentials in various
applications. Recently, in the efforts aiming at the coherent
control of single electrons or electron spins, the quantum transport
methods have been used to detect the quantized motion of electrons
in nanostructures.\cite{RMPspinQD,PRLMarcusTarucha} Besides the
average current, the current noises also contain useful information
about the quantum dynamics in
nanostructures.\cite{PhysReportNoise,PhysTodayShotNoise,PRLMarcusNoise}

Various theoretical approaches have been developed to treat the
quantum transport problem. The Landauer-B\"{u}tikker formula has
established the basic relationship between scattering amplitudes and
currents through nanostructures.\cite{Landauer,bookDatta1} The
non-equilibrium Green's function method provides a perturbation
scheme to deal with the many-body interaction effects in quantum
transport.\cite{bookQK,bookDatta2} In the past few years, approaches
based on notions in quantum optics were developed to study
time-dependent quantum transport processes in solid-state
structures. Most of these quantum optics approaches adopt the
density matrix formalism, with master equations or rate equations in
the Schr\"{o}dinger
picture.\cite{Nazarov,GurvitzPRBMasterEq,PRLBrandesA,PRLBrandesB,Kieblich,Aghassi,LiXinQiCurrent,LiXinQiNoise,DongBRate}
Very recently, quantum Langevin equation in the Heisenberg picture
was also used to establish the quantum rate equations to study the
transport problem. \cite{DongBLangevin}

\begin{figure}[tbp]
\includegraphics[bb=232 280 532 530, width=7 cm, clip]{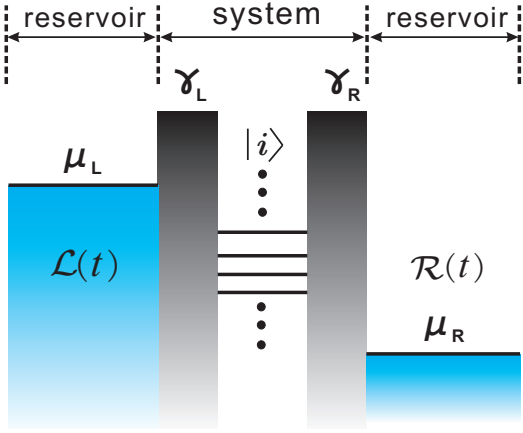}
\caption{{}(color online) Schematic illustration of transport
through a quantum dot. The whole system is divided into three parts,
the central system, and the left and right reservoirs. The central
system is characterized by eigen-states $|i\rangle$ with discrete
energies. Quantum noise operators $\mathcal{L}(t)$ and
$\mathcal{R}(t)$ are introduced to describe the reservoirs.}
\label{FigScheme}
\end{figure}

In this paper, instead of the master equations or the rate
equations, we will develop a quantum noise approach based on the
quantum Langevin equation to the quantum transport problem.
Essentially, we recognize that a general quantum transport problem
can be regarded as a system-plus-reservoir problem. In this sense,
the total system is divided into several sub-systems (see
Fig~\ref{FigScheme}). The \textit{central system} (\textit{system}
for short) is a nanostructure, such as a quantum dot or coupled
quantum dots. This subsystem contains several discrete
electronic energy levels, resembling an artificial atom. The
electrons in the leads, which have a continuous energy spectrum and
are kept in quasi-equilibrium, constitute the fermionic
\textit{reservoirs}. The electrons in the reservoirs can be treated
as free quasi-particles with the screened Coulomb interaction taken
into account as a renormalization of electron effective mass. The
central system and the reservoirs are coupled together to each other
through hopping across the barriers. With this observation, it is
natural to treat the quantum transport problem in the framework of
the quantum open system method, the quantum Langevin equation, a
standard approach in quantum optics to study cavity photon decay and
atom damping.

As compared to the application in quantum optics, the quantum
Langevin approach in the quantum transport problem has two features
to be singled out: (i) The reservoirs consist of electrons, which
are fermions while the baths in quantum optics are bosonic, and (ii)
when finite biases are applied between different leads, the
electronic reservoirs in different leads are in quasi-equilibrium
with different chemical potentials but do not stay in equilibrium
with each other. Our investigation in this paper will refine these
features. As illustrative applications of our approach, the resonant
transport through a single quantum dot is investigated for both the
single-level case and the Coulomb blockade case.

The quantum Langevin approach is a natural formalism to study the
noise spectroscopy of quantum dynamics in nanostructures,\cite{PRLMarcusNoise}
which is particularly interesting for small quantum systems where the
signals are often much weaker than the shot noises. When the
coupling between the leads and the nanostructures can be described
in the Markovian approximation, which is justified in large bias
cases, the quantum Langevin approach provides an exact treatment of
the interaction within the nanostructure. Furthermore, the quantum
Langevin equation establishes a fundamental relationship and analogy
between photon emission and electron tunneling processes, providing
new understanding of quantum transport phenomena with notions and
methods from quantum optics.

The paper is organized as follow. In Sec.~\ref{general}, we
introduce the basic concepts and the general formalism of the quantum
noise approach to treat the quantum transport problem. In
Sec.~\ref{application_I} and \ref{application_II}, we apply the
quantum noise approach to transport through a single quantum dot
containing a single level and double energy levels, respectively. In
Sec.~\ref{comparison}, we show the relations between our approach
and other quantum transport theories. We conclude and give an
outlook of our approach in Sec.\ref{conclusion}.

\section{General Formalism}
\label{general}
\subsection{Quantum Langevin equations for quantum transport}

\label{Sec_QLE}

In general, the quantum transport problem of nanostructures can be modeled
by the following Hamiltonian:%
\begin{equation}
H=H_{\text{sys}}\left(a_{i},a_{i}^{\dag }\right)+H_{\text{lead}}+H_{T},
\end{equation}%
where $H_{\text{sys}}$ describes the nanostructure, such as a
quantum dot, with multiple discrete energy levels. The leads, which
play the role of reservoirs, are described by the Hamiltonian
$H_{\text{lead}}$. The electron tunneling between the leads and the
nanostructure is included in $H_{T}$. For the two-lead case, the
leads Hamiltonian $H_{\text{lead}}$ and the tunneling Hamiltonian
$H_{T}$ can be written as
\begin{subequations}
\begin{eqnarray}
H_{\text{lead}} &=& \sum\limits_{k}\hbar
\omega_{k}^{(L)}b_{k}^{\dag}b_{k}+\sum\limits_{j}\hbar \omega
_{j}^{(R)}c_{j}^{\dag }c_{j}, \\
H_{T}& =&i\hbar \sum_{i,k}\xi _{ik} b_{k}^{\dag }a_{i} +i\hbar
\sum_{i,j}\zeta _{ij} c_{j}^{\dag }a_{i}+\mathrm{h.c.},
\end{eqnarray}
\end{subequations}
where $b_{k}$ and $c_{j}$ are the annihilation operators of the left
and right leads with continuous spectra $\hbar \omega _{k}^{(L)}$
and $\hbar \omega _{j}^{(R)},$ respectively. The tunneling is
characterized by the coefficients $\xi _{ik}$ and $\zeta _{ij}$.
Note that we have neglected the interaction in the leads as a common
approximation for a Fermi sea with the Coulomb interaction
effectively taken into the renormalized quasi-particle spectra.

Now, we consider the Heisenberg equations of motion of the system and
reservoir operators. For simplicity, we show equations of motion for the
simplest single level case, i.e. $a_{i}(t)= a(t)$, $\xi_{ik}=\xi_k$, $%
\zeta_{ij}=\zeta_j$and $H_{\text{sys}}=\hbar\omega_0 a^{\dag}a$. The
multi-level case will be discussed later in this paper. Straightforward
calculation gives
\begin{subequations}
\begin{eqnarray}
\dot{a}\left( t\right) &= &-i\omega_0 a - \sum_{k}\xi_{k}b_{k}-\sum_{j}\zeta
_{j}c_{j},  \label{GEsystemEOM} \\
\dot{b}_{k}\left( t\right) &=&-i\omega_{k}^{(L)}b_{k}+\xi _{k}a, \\
\dot{c}_{j}\left( t\right) &=&-i\omega_{j}^{(R)}c_{j} +\zeta _{j}a.
\end{eqnarray}
\end{subequations}
In the following, we try to eliminate the lead variables from the
equation of motion (\ref{GEsystemEOM}) of the system operator. To
this end, the formal solution for $b_{k}\left( t\right) $ is written
as
\begin{equation}
b_{k}\left( t\right) =e^{-i\omega _{k}t}b_{k}\left( 0\right) +\xi
_{k}\int_{0}^{t}dt^{\prime }\left[ e^{-i\omega _{k}\left( t-t^{\prime
}\right) }a\left( t^{\prime }\right) \right] .
\end{equation}%
With this formal solution, the following relation is obtained%
\begin{equation}
\sum\limits_{k}\xi _{k}b_{k}\left( t\right)
=-\mathcal{L}_{\text{in}}\left( t\right) +\frac{\gamma
_{L}}{2}a\left( t\right),  \label{GESummationRelation}
\end{equation}
where the \textit{input} noise operator $\mathcal{L}_{\text{%
in}}\left( t\right)$ due to the left lead is defined as
\begin{equation}
\mathcal{L}_{\text{in}}\left( t\right) =-\sum_{k}\xi _{k}e^{-i\omega
_{k}^{(L)}t}b_{k}\left( 0\right) .
\end{equation}
The damping term in Eq.~(\ref{GESummationRelation}) arises from the
Markovian approximation\cite{bookGardiner} under the continuous
limit
\begin{eqnarray}
&&\sum_{k}\xi _{k}^{2}e^{-i\omega _{k}^{(L)}\left( t-t^{\prime
}\right) }
\nonumber
\\
&=&\int d\omega _{k}\left[ D\left( \omega _{k}\right) \xi ^{2}\left( \omega
_{k}\right) e^{-i\omega _{k}^{(L)}\left( t-t^{\prime }\right) }\right] \notag\\
&=&\gamma _{L}\delta \left( t-t^{\prime }\right),
\end{eqnarray}
where $D(\omega_k)$ is the density of the states of the lead. Here
we have assumed that $D(\omega_k)\xi^2(\omega_k)$ is flat around the
frequency $\omega_0$ and
\begin{equation}
\gamma _{L}=2\pi D\left( \omega _{0}\right) \xi ^{2}\left( \omega
_{0}\right),
\end{equation}%
is widely used as the tunneling rate in nanostructure quantum transport problems.

Similarly, for the right lead
\begin{equation}
\sum\limits_{j}\zeta _{j}c_{j}\left( t\right) =-\mathcal{R}_{\text{in}%
}\left( t\right) +\frac{\gamma _{R}}{2}a\left( t\right), \label{GESummationRelationRight}
\end{equation}
where the noise operator of the right lead $\mathcal{R}\left(
t\right) $ is
defined as%
\begin{equation}
\mathcal{R}_{\text{in}}\left( t\right)
=-\sum_{j}\zeta_{j}e^{-i\omega _{j}^{(R)}t}c_{j}\left( 0\right) ,
\end{equation}%
and $\gamma_R$ is the tunneling rate to the right lead. Using
Eqs.~(\ref{GESummationRelation}) and
(\ref{GESummationRelationRight}), we obtain the quantum Langevin
equation for the system operator $a(t)$:
\begin{eqnarray}
\dot{a}\left( t\right) =-i\omega_0 a(t) -\frac{\gamma _{L}+\gamma _{R}}{2}%
a\left( t\right) +\mathcal{L}_{\text{in}}\left( t\right) +\mathcal{R}_{\text{%
in}}\left( t\right).  \label{GeneralLangevin}
\end{eqnarray}%
Similar to the cases in quantum optics, the two electronic leads,
which play the role of fermionic reservoirs, induce the damping and
the fluctuations through the noise operators.

\subsection{Projection operator formalism for interacting systems}
\label{Sec_projection}%
For an interacting system, the complexity of the quantum Langevin
equations arises from the evolution induced by the system
Hamiltonian $H_{\text{sys}}$. To deal with such complexity,
we introduce the projection operators of the interacting system in
this subsection.

Though there may be interaction between the electrons in the system
Hamiltonian $H_{\text{sys}}$, the artificial atom can always be
considered as consisting of a few discrete many-body energy levels.
In other words, we can diagonalize the system Hamiltonian
$H_{\text{sys}} $ as
\begin{equation}
H_{\text{sys}}\left(a_{i},a_{i}^{\dag}\right) =\sum\limits_{i=1}^{N}\hslash
\omega _{i}\left\vert i\rangle \langle
i\right\vert=\sum\limits_{i=1}^{N}\hslash \omega _{i}\sigma _{ii},
\end{equation}%
where $\sigma _{ij}=\left\vert i\rangle \langle j\right\vert $ is the
projection operator from $\left\vert j\right\rangle $ to $\left\vert
i\right\rangle $, with $\left\vert i\right\rangle$ being the eigen-state of $%
H_{\text{sys}}$ of energy $\hbar \omega _{i}$.

In general cases, determined by the system Hamiltonian $H_{\text{sys}}$, the
fermionic operators $a_{i}$ and $a_{i}^{\dag }$ can be written in terms of the projection operators as%
\begin{equation}
a_{i}=\sum_{k,l}T_{kl}^{(i)}\sigma _{kl}\text{ and }a_{i}^{\dag }=\sum_{k,l}%
\tilde{T}_{kl}^{(i)}\sigma _{kl},
\end{equation}%
where $T^{(i)}$ and $\tilde{T}^{(i)}$ are $N\times N$ matrices associated
with the fermionic operators $a_{i}$ and $a_{i}^{\dag }$, and $\tilde{T}%
^{(i)}\equiv\left(T^{(i)}\right)^{\dagger}$.

The commutative relation between the projection operators $\sigma
_{ij}$ can be calculated with $\sigma _{ij}\sigma _{kl}=\sigma
_{il}\delta _{jk} $. In the following calculations we also need to
identify the commutative relation between the projection operators
$\sigma _{ij}$ and the reservoir operators, i.e. $b_{k}(b_{k}^{\dag
})$ and $c_{j}(c_{j}^{\dag })$. Note that the eigen-state
$\left\vert i\right\rangle $ is also an eigen-state of the electron
number operator in the quantum dot $\hat{N}=\sum_{i}a_{i}^{\dag
}a_{i}$, i.e. $\hat{N}\left\vert i\right\rangle =N_{i}\left\vert
i\right\rangle $. Thus, the projection operator $\sigma _{ij}$
corresponds to a definite electron number change $N_{i}-N_{j}$,
which is either odd or even. Consequently, $\sigma _{ij}$ and
$b_{k}(b_{k}^{\dag })$ have the following (anti-)commutative
relation:
\begin{equation}
\left[ \sigma _{ij},b_{k}\right] _{g_{ij}}\equiv
\sigma_{ij}b_{k}+g_{ij}b_{k}\sigma _{ij}=0,
\end{equation}%
where the factor
\begin{equation}\label{parity}
g_{ij}=\left\{%
\begin{array}{cc}
1, & \text{for }N_{i}-N_{j}=\text{odd} \\
-1, & \text{for }N_{i}-N_{j}=\text{even}%
\end{array}%
\right. .
\end{equation}
The Heisenberg equation for the $\sigma _{ij}$ is
\begin{eqnarray}
\dot{\sigma}_{ij}\left( t\right) &=&\frac{i}{\hbar }\left[ H_{\text{sys}},
\sigma _{ij}\right] +\frac{i}{\hbar }\left[ H_{T},\sigma _{ij}\right] \notag \\
&=&-i\Delta _{ij}\sigma _{ij}\left( t\right) \notag\\
&&-\sum_{\alpha,k,m,n}\left[\left(\xi _{\alpha k}T_{mn}^{\left( \alpha\right)
}b_{k}^{\dag }\sigma _{mn}-\text{h.c.}\right),\sigma_{ij}\right]\notag\\
&&-\sum_{\alpha,j,m,n}\left[\left(\zeta _{\alpha j}T_{mn}^{\left(
\alpha\right) }c_{j}^{\dag }\sigma _{mn}-\text{h.c.}\right),\sigma
_{ij}\right].
\end{eqnarray}
With the help of the definition of noise operators and the first
Markovian approximation, we obtain the quantum Langevin equation for
the projection
operator $\sigma _{ij}$
\begin{eqnarray}
\dot{\sigma}_{i j }\left( t\right)
 &= &-i\Delta _{i j}\sigma _{i j }\left(t\right) \nonumber   \\
&& -\frac{\gamma _{L}}{2}\sum\limits_{m ,m ^{\prime }}D_{m m ^{\prime }}^{i
j}\sigma _{m m ^{\prime }}\left( t\right) +\left(\gamma _{L}\rightarrow
\gamma _{R}\right)  \notag \\
&&+\sum\limits_{m ,m ^{\prime }}C_{m m ^{\prime }}^{i j}\mathcal{L}^{\dag
}_{\text{in}}\left( t\right) \sigma _{m m ^{\prime }}\left(t\right) +\left(\mathcal{L}%
^{\dag }_{\text{in}}\rightarrow {\mathcal{R}}^{\dag}_{\text{in}}\right)  \notag \\
& &+\sum\limits_{m ,m ^{\prime }}\tilde{C}_{m m ^{\prime }}^{ij }\mathcal{L}_{\text{in}}%
\left( t\right) \sigma _{m m ^{\prime }}\left(t\right) +\left(\mathcal{L}_{\text{in}}%
\rightarrow \mathcal{R}_{\text{in}}\right),
\label{GECorrelationPOp}
\end{eqnarray}%
where $\Delta _{i j }=\omega _{j }-\omega_{i}$, and the
coefficients $D_{m m ^{\prime }}^{i j }$, $C_{m m ^{\prime }}^{i j }$, and $%
\tilde{C}_{m m ^{\prime }}^{i j }$ are defined as follows
\begin{subequations}
\begin{eqnarray}
D_{m m ^{\prime }}^{i j } &= & A_{m m ^{\prime }}^{i j }+\tilde{A}_{m
m^{\prime }}^{i j }+B_{m m ^{\prime }}^{i j }+\tilde{B}_{m m ^{\prime}}^{i j
}, \\
C_{m m ^{\prime }}^{i j }&=& \sum\limits_{\alpha}\left( T_{m i}^{\left(
\alpha\right) }\delta _{m ^{\prime }j }+g_{i j }T_{j m^{\prime }}^{\left(
\alpha\right) }\delta _{m i }\right) , \\
\tilde{C}_{m m ^{\prime }}^{i j }&=&\sum\limits_{\alpha}\left(\tilde{T}_{m i
}^{\left( \alpha\right) }\delta _{m ^{\prime }j}+g_{i j }\tilde{T}_{j m
^{\prime }}^{\left( \alpha\right) }\delta_{m i }\right),
\end{eqnarray}
\end{subequations}
with
\begin{subequations}
\begin{eqnarray}
A_{m m ^{\prime }}^{i j }&=&\sum\limits_{\alpha,\alpha^{\prime }}g_{i j }%
\tilde{T}_{m i }^{( \alpha^{\prime }) }T_{j m ^{\prime}}^{\left(
\alpha\right) }, \\
\tilde{A}_{m m ^{\prime }}^{i j }&=&\sum\limits_{\alpha,\alpha^{\prime
}}g_{i j }T_{m i }^{( \alpha^{\prime }) }\tilde{T}_{j m ^{\prime}}^{\left(
\alpha\right) }, \\
B_{m m ^{\prime }}^{i j}&=&\delta _{m ^{\prime }j
}\sum\limits_{\nu,\alpha,\alpha^{\prime }}\tilde{T}_{m\nu }^{(
\alpha^{\prime }) }T_{\nu i }^{\left( \alpha\right) }, \\
\tilde{B}_{m m ^{\prime }}^{i j }&=&\delta _{m ^{\prime
}j}\sum\limits_{\nu,\alpha,\alpha^{\prime }}T_{m \nu }^{( \alpha^{\prime })}%
\tilde{T}_{\nu i }^{\left( \alpha\right) }.
\end{eqnarray}
\end{subequations}

In principle, the quantum Langevin equation for the system operators
is equivalent to a quantum stochastic equation if we introduce the
quantum Wiener process,\cite{PapaerGardiner1} and the properties of
the their solution can be discussed by defining the quantum
stochastic integration.\cite{PapaerGardiner1} Thus, we point out
that the quantum transport problem provides an experimentally
accessible proving ground for the quantum stochastic theory. Instead
of further discussing the mathematical properties of
Eq.~(\ref{GECorrelationPOp}), in this paper we will focus, through
concrete models, on how to derive the observable quantities in
quantum transport.

\subsection{Boundary Relation and Causality}\label{Sec:causality}
Besides the \textit{input} noise operators
$\mathcal{L}_{\text{in}}\left( t\right)$ and
$\mathcal{R}_{\text{in}}\left( t\right)$, the \textit{output} noise
operators\cite{bookDFWalls,PapaerGardiner1} can be defined as
\begin{subequations}
\begin{eqnarray}
\mathcal{L}_{\text{out}}\left( t\right) &=& -\sum_{k}\xi
_{k}e^{-i\omega
_{k}^{(L)}(t-t_{\text{f}})}b_{k}\left( t_{\text{f}}\right), \\
\mathcal{R}_{\text{out}}\left( t\right) &=& -\sum_{j}\zeta
_{j}e^{-i\omega _{j}^{(R)}(t-t_{\text{f}})}c_{j}\left(
t_{\text{f}}\right),
\end{eqnarray}
\end{subequations}
where $t_{\text{f}}$ is a time in the remote future. Similar to
Eqs.~(\ref{GESummationRelation}) and
(\ref{GESummationRelationRight}), the first Markovian approximation
gives the following relations
\begin{equation}
\sum\limits_{k}\xi _{k}b_{k}\left( t\right)
=-\mathcal{L}_{\text{out}}\left( t\right) -\frac{\gamma
_{L}}{2}a\left( t\right),  \label{GESummationRelationOUT}
\end{equation}

According to Eqs.~(\ref{GESummationRelation}) and
(\ref{GESummationRelationOUT}), the ``boundary relation'' between
the noise operators and the system operator is \cite{bookDFWalls}
\begin{equation}
\mathcal{L}_{\text{in}}\left(
t\right)-\mathcal{L}_{\text{out}}\left(
t\right)=\gamma_{L}a(t).\label{boundaryRelation}
\end{equation}
Similarly, for the right lead
\begin{equation}\label{boundaryRelation1}
\mathcal{R}_{\text{in}}\left(
t\right)-\mathcal{R}_{\text{out}}\left( t\right)=\gamma_{R}a(t).
\end{equation}

According to the quantum Langevin equation (\ref{GeneralLangevin}),
the fermionic system operator $d(t)\in\{a(t),a^{\dagger}(t)\}$ at
time $t$ only depends on the input noise operators at time $t'<t$.
As a result, in the Markovian limit, the causality relation reads
\cite{PapaerGardiner1,PapaerGardiner3}
\begin{equation}
\left[\mathcal{L}_{\text{in}}\left( t'\right),d(t)\right]_{+}=0,
\text{ for } t'>t.\label{causality1}
\end{equation}
For the similar reason, the system operator at $t$ is independent of
the output noise operators at time $t'<t$
\begin{equation}
\left[\mathcal{L}_{\text{out}}\left( t'\right),d(t)\right]_{+}=0,
\text{ for } t'<t.\label{causality2}
\end{equation}
According to
Eqs.~(\ref{boundaryRelation}$\thicksim$\ref{causality2}), the
anti-commutators between noise and system operator are converted to
those between system operators\cite{PapaerGardiner1,PapaerGardiner3}
\begin{subequations}\label{causalities}
\begin{eqnarray}
\left[\mathcal{L}_{\text{in}}\left(
t'\right),d(t)\right]_{+}&=&\gamma_{L}\theta(t-t')[a(t'),d(t)]_{+},\label{causality3}\\
\left[\mathcal{R}_{\text{in}}\left(
t'\right),d(t)\right]_{+}&=&\gamma_{R}\theta(t-t')[a(t'),d(t)]_{+},\label{causality5}
\end{eqnarray}
\end{subequations}
where step function $\theta(t)$ is defined as
\begin{equation}
\theta \left( t\right) =\left\{
\begin{array}{c}
1, \\
\frac{1}{2}, \\
0,%
\end{array}%
\begin{array}{c}
t>0 \\
t=0 \\
t<0%
\end{array}%
\right..
\end{equation}
For the multi-level case, this causality relation
Eq.~(\ref{causalities}) can be generalized to the system projection
operators $\sigma_{ij}$, i.e.
\begin{subequations}\label{causalitiesPOp}
\begin{eqnarray}
\left[\mathcal{L}_{\text{in}}\left(
t'\right),\sigma_{ij}(t)\right]_{\pm}&=&\gamma_{L}\theta(t-t')[a(t'),\sigma_{ij}(t)]_{\pm},\label{causalityPOp3}\\
\left[\mathcal{R}_{\text{in}}\left(
t'\right),\sigma_{ij}(t)\right]_{\pm}&=&\gamma_{R}\theta(t-t')[a(t'),\sigma_{ij}(t)]_{\pm}.\label{causality6}
\end{eqnarray}
\end{subequations}
The choice of the commutative and anti-commutative relation in
Eq.~(\ref{causalitiesPOp}) is determined by the parity of the
electron number change, see Eq.~(\ref{parity}).

In the following, to simplify the notation, we will omit the
subscript ``in'' of the input noise operators, unless stated
otherwise.

\subsection{Current and current noise}
For the quantum transport problem, we are interested in the average
current and the current noise spectra. In this subsection, we will
give the expressions of such quantities in terms of the noise
operators.

We consider the current through the right lead for example. For
simplicity, let us first study the single level case. The formula
for the multi-level case with Coulomb blockade will be discussed
later. The current operator can be defined as the changing rate of
the electron number on the right lead, i.e.
\begin{subequations}
\begin{eqnarray}
\hat{I}_{R} &=&\frac{d}{dt}\hat{N}_{R} =\sum_{j}\zeta _{j}c_{j}^{\dag }a+%
\mathrm{h.c.}  \label{CurrentOP} \\
&=&\gamma _{R}a^{\dag }\left( t\right) a\left( t\right)
-\mathcal{R}^{\dag }\left( t\right) a\left( t\right) -a^{\dag}\left(
t\right) \mathcal{R}\left( t\right).   \label{CurrentOPDef}
\end{eqnarray}
\end{subequations}
The second line is obtained by noticing the relations in Eq.~(\ref%
{GESummationRelation}) and Eq.~(\ref{GESummationRelationRight}). We
point
out that the current operator can be divided into two parts: (i) the \textit{%
damping part} $\gamma _{R}a^{\dag }\left( t\right) a\left( t\right)
$, which is proportional to the level occupation and the escaping
rate $\gamma _{R}$; and (ii) the \textit{fluctuation part} (the last
two terms), which is
induced by the noise operators $\mathcal{R}\left( t\right) $ and $\mathcal{R}%
^{\dag }\left( t\right) $.

For the average current, we take the average of the current operator $\hat{I}%
_{R}$ over the thermal states of the leads
\begin{equation}
\langle \hat{I}_{R}\rangle =\gamma _{R}\langle a^{\dag }\left(
t\right) a\left( t\right) \rangle -\langle \mathcal{R}^{\dag }\left(
t\right) a\left( t\right) \rangle -\langle a^{\dag }\left( t\right)
\mathcal{R}\left( t\right) \rangle.
\end{equation}
And for the current noise, we first calculate the current-current
correlation function%
\begin{equation}
g^{\left( 2\right) }\left( \tau \right) =\lim_{t\rightarrow +\infty }Re\left[%
\langle \hat{I}_{R}\left( t\right) \hat{I}_{R}\left( t+\tau \right) \rangle%
\right] -\langle \hat{I}_{R}\rangle ^{2}
\end{equation}%
At steady state, its Fourier transformation gives the current noise spectrum%
\cite{bookNoise}
\begin{equation}
S\left( \omega \right) =4\int_{0}^{\infty }g^{\left( 2\right)
}\left( \tau \right) \cos \left( \omega \tau \right) d\tau .
\end{equation}
To calculate the correlation $\langle \hat{I}_{R}\left( t\right) \hat{I%
}_{R}\left( t+\tau \right) \rangle$ in $g^{\left( 2\right) }\left(
\tau \right)$, by the definition of $\hat{I}_{R}$ in
Eq.~(\ref{CurrentOPDef}), one need to calculate the two time
correlation such as
\begin{subequations}
\label{GECorrelation}
\begin{eqnarray}
&& \langle a^{\dag }\left( t\right) a\left( t\right) a^{\dag}\left(
t+\tau
\right) a\left( t+\tau \right)\rangle, \\
&& \langle a^{\dag }\left( t\right) \mathcal{R}\left( t\right)
a^{\dag}\left( t+\tau \right) a\left( t+\tau\right)\rangle, \\
&& \langle a^{\dag }\left( t\right) \mathcal{R}\left( t\right)\mathcal{R}%
^{\dag }\left( t+\tau \right) a\left( t+\tau \right)\rangle.
\end{eqnarray}
\end{subequations}
In Sect.~\ref{application_I}, we will show that the fluctuation part
in the current operator does not contribute to the average current,
so the average current $\langle\hat{I}_{R}\rangle=\gamma _{R}\langle
a^{\dag }\left( t\right) a\left( t\right)\rangle$ is held. But the
fluctuation terms will contribute to the current noise through the
correlations in Eq.~(\ref{GECorrelation}).

\section{Application I: Single level transport}
\label{application_I}%

In this section, the general quantum Langevin formula is applied to
the resonant transport through a quantum dot. As the first example,
we consider a model in which only one single energy level in the
quantum dot is relevant. The system Hamiltonian reads
\begin{equation}
H_{\text{sys}}=\hbar \omega _{0}a^{\dag }a.
\end{equation}%
We consider the large bias condition and assume that the
single-particle energy level with energy $\hbar \omega _{0}$ is well
within the bias window, i.e. $\mu _{L}-\omega _{0}$, $\omega_0-\mu
_{R} \gg \gamma_L$, $\gamma_R$, for $\mu _{L/R}$ being the chemical
potentials of the left/right leads. According to the
discussion in Sec.~\ref{Sec_QLE}, the quantum Langevin equation reads%
\begin{equation}
\dot{\tilde{a}}\left( t\right) =-\frac{\gamma _{L}+\gamma _{R}}{2}\tilde{a}%
\left( t\right) +\tilde{\mathcal{L}}\left( t\right) +\tilde{\mathcal{R}}%
\left( t\right) ,  \label{singellevelLagevinEq}
\end{equation}%
where $\tilde{a}\left( t\right) =a\left( t\right) e^{i\omega _{0}t}$,
$\tilde{\mathcal{L}}\left(
t\right) =e^{i\omega _{0}t}\mathcal{L}\left( t\right) $, and $\tilde{\mathcal{R}}%
\left( t\right) =e^{i\omega _{0}t}\mathcal{R}\left( t\right) $ are defined
in the rotating reference frame to single out the slow-varying dynamics. In
the white-noise limit, the correlation between the noise operators
can be written as (see Appendix~\ref{appendix_A})
\begin{subequations}
\begin{eqnarray}
\langle \tilde{\mathcal{L}}^{\dag }\left( t\right) \tilde{\mathcal{L}}\left(
t^{\prime }\right) \rangle &=&\gamma _{L}\delta \left( t-t^{\prime }\right) ,
\\
\langle \tilde{\mathcal{R}}\left( t\right) \tilde{\mathcal{R}}^{\dag }\left(
t^{\prime }\right) \rangle &=&\gamma _{R}\delta \left( t-t^{\prime }\right) ,
\\
\langle \tilde{\mathcal{L}}\left( t\right) \tilde{\mathcal{L}}^{\dag }\left(
t^{\prime }\right) \rangle &=&
\langle \tilde{\mathcal{R}}^{\dag }\left( t\right) \tilde{\mathcal{R}}\left( t^{\prime }\right)
\rangle =0.
\end{eqnarray}%
\end{subequations}
Using these relations, we calculate average current and current
noise.

\subsection{Average current}

From Eq.~(\ref{singellevelLagevinEq}), the system operator
$\tilde{a}\left( t\right) $ in terms of the noise operators is
\begin{eqnarray}
\tilde{a}\left( t\right)  &=&e^{-\frac{\Gamma }{2}t}\tilde{a}\left( 0\right)
+\int_{0}^{t}e^{-\frac{\Gamma }{2}\left( t-t^{\prime }\right) }\mathcal{%
\tilde{L}}\left( t^{\prime }\right) dt^{\prime }   \notag \\
&&+\int_{0}^{t}e^{-\frac{\Gamma }{2}\left( t-t^{\prime }\right) }\tilde{%
\mathcal{R}}\left( t^{\prime }\right) dt^{\prime },
\label{SinglelevelSolution}
\end{eqnarray}%
where $\Gamma =\gamma _{L}+\gamma _{R}$. Multiplying the noise
operator $\mathcal{\tilde{L}}^{\dag }\left( t\right)$ on both sides of Eq.~(\ref{SinglelevelSolution}), we have%
\begin{eqnarray}
\langle \mathcal{\tilde{L}}^{\dag }\left( t\right) \tilde{a}\left( t\right)
\rangle  &=&e^{-\frac{\Gamma }{2}t}\langle \mathcal{\tilde{L}}^{\dag }\left(
t\right) \tilde{a}\left( 0\right) \rangle    \notag \\
&&+\int_{0}^{t}e^{-\frac{\Gamma }{2}\left( t-t^{\prime }\right) }\langle
\mathcal{\tilde{L}}^{\dag }\left( t\right) \mathcal{\tilde{L}}\left(
t^{\prime }\right) \rangle dt^{\prime }  \notag \\
&=&\int_{0}^{t}dt^{\prime }\left[ e^{-\frac{\Gamma }{2}\left( t-t^{\prime
}\right) }\gamma _{L}\delta \left( t-t^{\prime }\right) \right]   \notag \\
&=&\frac{\gamma _{L}}{2}.
\end{eqnarray}%
Here we have assumed that at initial time $t=0$, the system and the
reservoir are independent, i.e., $\langle \mathcal{\tilde{L}}^{\dag }\left(
t\right) \tilde{a}\left( 0\right) \rangle =\langle \mathcal{\tilde{L}}^{\dag
}\left( t\right) \rangle \langle \tilde{a}\left( 0\right) \rangle =0$.
Similarly, we obtain
\begin{equation}
\langle \tilde{a}^{\dag }\left( t\right) \mathcal{\tilde{L}}\left( t\right)
\rangle =\frac{\gamma _{L}}{2},
\end{equation}%
and%
\begin{equation}
\langle \tilde{a}^{\dag }\left( t\right) \tilde{\mathcal{R}}\left( t\right)
\rangle =\langle \tilde{\mathcal{R}}^{\dag }\left( t\right) \tilde{a}\left(
t\right) \rangle =0.  \label{singlelevelcorr}
\end{equation}%
Thus, according to Eq.~(\ref{CurrentOPDef}), the fluctuation part of
the current operator does not contribute to the average current, and
the average current becomes
\begin{equation}
\langle I_{R}\rangle =\gamma _{R}\langle \tilde{a}^{\dag }\tilde{a}\rangle .
\end{equation}%
In order to determine the mean occupation number $\langle \tilde{a}^{\dag }%
\tilde{a}\rangle $, we use the equation of motion
\begin{eqnarray}
\frac{d}{dt}\tilde{a}^{\dag }\tilde{a} &=&\dot{a}^{\dag }\tilde{a}+\tilde{a}%
^{\dag }\dot{a}   \notag \\
&=&-\Gamma \tilde{a}^{\dag }\tilde{a}+\tilde{a}^{\dag }\left( t\right)
\mathcal{\tilde{L}}\left( t\right) +\mathcal{\tilde{L}}^{\dag }\left(
t\right) \tilde{a}\left( t\right)   \notag \\
&&+\tilde{a}^{\dag }\left( t\right) \tilde{\mathcal{R}}\left( t\right) +%
\tilde{\mathcal{R}}^{\dag }\left( t\right) \tilde{a}\left( t\right).
\end{eqnarray}%
The ensemble average leads to%
\begin{equation}
\frac{d}{dt}\langle \tilde{a}^{\dag }\tilde{a}\rangle =-\Gamma \langle
\tilde{a}^{\dag }\tilde{a}\rangle +\gamma _{L}.  \label{SinglelevelAverEq}
\end{equation}%
Thus, the averaged population in the quantum dot is
\begin{equation}
\langle \tilde{a}^{\dag }\tilde{a}\rangle =\frac{\gamma _{L}}{\gamma
_{L}+\gamma _{R}}-\frac{\gamma _{L}}{\gamma _{L}+\gamma _{R}}e^{-\left(
\gamma _{L}+\gamma _{R}\right) t}.
\end{equation}%
As a result, the average current at steady state for $t\rightarrow +\infty $
is%
\begin{equation}
\langle \hat{I}_{R}\rangle _{ss}=\frac{\gamma _{L}\gamma _{R}}{\gamma
_{L}+\gamma _{R}},
\end{equation}%
which is the well-known result for the resonant tunneling
transport.\cite{bookDatta2,GurvitzPRBMasterEq}

\subsection{Current noise}
To investigate the current noise, we calculate
the current-current correlation $\langle \hat{I}_{R}\left( t\right) \hat{I}%
_{R}\left( t+\tau \right) \rangle $. With the definition of the current operator in Eq.~(%
\ref{CurrentOPDef}), the noise contains typically two-time
correlations like
\begin{equation}
\langle a^{\dag }\left( t\right) a\left( t\right) a^{\dag }\left( t+\tau
\right) a\left( t+\tau \right) \rangle \equiv \langle \hat{n}\left( t\right)
\hat{n}\left( t+\tau \right) \rangle ,
\end{equation}%
and%
\begin{equation}
\langle a^{\dag }\left( t\right) \tilde{\mathcal{R}}\left( t\right) \tilde{%
\mathcal{R}}^{\dag }\left( t+\tau \right) a\left( t+\tau \right) \rangle .
\end{equation}%
We will discuss such correlations one by one.

Noticing that the electron number correlation function $\langle \hat{n}%
\left( t\right) \hat{n}\left( t+\tau\right)\rangle $ contains only the
system operators, we use the quantum regression theorem\cite{bookScully} and Eq.~(\ref{SinglelevelAverEq})
and obtain
\begin{equation}
\frac{d}{d\tau }\langle \hat{n}\left( t\right) \hat{n}\left( t+\tau \right)
\rangle =-\Gamma \langle \hat{n}\left( t\right) \hat{n}\left( t+\tau \right)
\rangle +\gamma _{L}\langle \hat{n}\left( t\right) \rangle .
\end{equation}%
This equation, together with the initial condition with respect to
$\tau $, i.e. for $\tau =0$, $\langle \hat{n}\left( t\right)
\hat{n}\left( t+\tau \right) \rangle =\langle \hat{n}\left( t\right)
\hat{n}\left( t\right) \rangle =\langle \hat{n}\left( t\right)
\rangle $, determines the occupation number fluctuation in the
quantum dot. The steady state correlation is
\begin{eqnarray}
&& \lim_{t\rightarrow +\infty }\langle \hat{n}\left( t\right) \hat{n}\left(
t+\tau \right) \rangle \nonumber  \\
&=&\frac{\gamma _{L}^{2}}{\left( \gamma _{L}+\gamma _{R}\right) ^{2}}+\frac{%
\gamma _{L}\gamma _{R}}{\left( \gamma _{L}+\gamma _{R}\right) ^{2}}%
e^{-\left( \gamma _{L}+\gamma _{R}\right) \tau }.
\label{numnumCorr}
\end{eqnarray}
The other terms contain the correlations between the system and noise
operators. Taking $\langle \tilde{a}^{\dag }\left( t\right) \tilde{\mathcal{R%
}}\left( t\right) \tilde{\mathcal{R}}^{\dag }\left( t+\tau \right) \tilde{a}%
\left( t+\tau \right) \rangle $ for example, according to Eq.~(\ref%
{SinglelevelSolution}), we have
\begin{eqnarray}
&&\langle \tilde{a}^{\dag }\left( t\right) \tilde{\mathcal{R}}\left(
t\right) \tilde{\mathcal{R}}^{\dag }\left( t+\tau \right) \tilde{a}\left(
t+\tau \right) \rangle   \notag \\
&=&\int_{0}^{t}dt_{1}\int_{0}^{t+\tau }dt_{2}e^{-\frac{\Gamma }{2}\left(
t-t_{1}\right) -\frac{\Gamma }{2}\left( t+\tau -t_{2}\right)
}G(t_{1},t,t+\tau ,t_{2}),
\end{eqnarray}%
where the four-time noise correlation is defined as
\begin{eqnarray}
G(t_{1},t_{2},t_{3},t_{4})   =\big\langle\tilde{\mathcal{L}}^{\dag }\left( t_{1}\right) \tilde{\mathcal{
R}}\left( t_{2}\right) \tilde{\mathcal{R}}^{\dag }\left( t_{3}\right) \tilde{\mathcal{L}}%
\left( t_{4}\right) \big\rangle.
\end{eqnarray}%
According to the independent noise assumption and the white-noise
approximation,
\begin{equation}
G(t_{1},t_{2},t_{3},t_{4})=\gamma _{L}\gamma _{R}\delta \left(
t_{1}-t_{4}\right) \delta \left( t_{2}-t_{3}\right) .
\end{equation}%
Thus, we have
\begin{eqnarray}
&&\langle \tilde{a}^{\dag }\left( t\right) \tilde{\mathcal{R}}\left(
t\right) \tilde{\mathcal{R}}^{\dag }\left( t+\tau \right) \tilde{a}\left(
t+\tau \right) \rangle   \notag \\
&=&\int_{0}^{t}dt_{1}\int_{0}^{t+\tau }dt_{2}\left[ e^{-\frac{\Gamma }{2}%
\left( 2t+\tau -t_{1}-t_{2}\right) }\delta \left( t_{1}-t_{2}\right) \delta
\left( \tau \right) \right]  \notag \\
&=&\frac{\gamma _{L}\gamma _{R}}{\gamma _{L}+\gamma _{R}}e^{-\frac{\gamma
_{L}+\gamma _{R}}{2}\tau }\delta \left( \tau \right) ,\text{ \ for }%
t\rightarrow +\infty .  \label{flucfluc}
\end{eqnarray}%
Similarly,
\begin{eqnarray}
&&\langle \tilde{a}^{\dag }\left( t\right) \mathcal{\tilde{R}}\left(
t\right) \tilde{a}^{\dag }\left( t+\tau \right) \tilde{a}\left( t+\tau
\right) \rangle   \notag \\
&=&\frac{\gamma _{L}\gamma _{R}}{\gamma _{L}+\gamma _{R}}e^{-(\gamma
_{L}+\gamma _{R})\tau },\text{ \ for }t\rightarrow +\infty .
\label{numfluct}
\end{eqnarray}%
It can be checked that all the other terms in the current-current
correlation function vanish. Consequently, the current-current
correlation function is%
\begin{equation}
g^{\left( 2\right) }\left( \tau \right) =-\frac{\gamma _{L}^{2}\gamma
_{R}^{2}}{\left( \gamma _{L}+\gamma _{R}\right) ^{2}}e^{-\Gamma \tau }+\frac{%
\gamma _{L}\gamma _{R}}{\gamma _{L}+\gamma _{R}}e^{-\frac{\Gamma }{2}\tau
}\delta \left( \tau \right) ,
\end{equation}%
and its Fourier transformation gives the current noise spectra%
\begin{equation}\label{singleCurrentNoise}
S\left( \omega \right) =2e\langle \hat{I}_{R}\rangle
_{ss}\frac{\gamma _{L}^{2}+\gamma _{R}^{2}+\omega ^{2}}{\left(
\gamma _{L}+\gamma _{R}\right) ^{2}+\omega ^{2}}.
\end{equation}%
This result accords with the ones derived from other
approaches,\cite {LiXinQiNoise} and shows that the presence of the
single level quantum dot
suppresses the zero-frequency current noise to half of the Poisson value $%
S_{P}=2e\langle I_{R}\rangle _{ss}$ in the case $\gamma _{L}=\gamma
_{R}$.

It is worth to emphasize that, clearly shown in our quantum noise
approach, although the fluctuation part [see
Eq.~(\ref{CurrentOPDef})] of the current operator does not
contribute to the average current, it does the current noise.
According to our approach, the current-current correlation
originates from three different kinds of sources: (i) the on-site
number-number correlation [Eq.~(\ref{numnumCorr})], which always
contributes a positive correlation, (ii) the correlation between the
fluctuation terms Eq.~(\ref{flucfluc}), which induces a white-noise
correlation, and (iii) and
the correlation between the on-site number and the fluctuation term Eq.~(\ref%
{numfluct}), which always provides a negative correlation. This
classification of current-current correlation is also valid in the
interacting case, as will be discussed below.

\section{Application II: Coulomb Blockade}

\label{application_II}

\subsection{Average current}

Now we apply the general theory to the Coulomb blockade case. For
simplicity, we assume that only one single orbital level in the
quantum dot is relevant (i.e., within the energy range of interest).
The system Hamiltonian reads
\begin{equation}
H_{\text{sys}}\left( a_{i},a_{i}^{\dag }\right) =\hbar \omega
_{\uparrow }a_{\uparrow }^{\dag }a_{\uparrow }+\hbar \omega
_{\downarrow }a_{\downarrow }^{\dag }a_{\downarrow }+Ua_{\uparrow
}^{\dag }a_{\uparrow }a_{\downarrow }^{\dag }a_{\downarrow },
\end{equation}%
where $\hbar \omega _{\uparrow ,\downarrow }$ are the single
electron energy for spin-up and spin-down electrons in the quantum
dot, and $U$ is the Coulomb interaction strength between two
electrons. In this paper, we consider the large $U$ limit, i.e.
$\hbar \omega _{\uparrow }+U,\hbar \omega _{\downarrow }+U\gg \mu
_{L}\gg \hbar \omega _{\uparrow },\hbar \omega _{\downarrow }\gg \mu
_{R}$.

\begin{figure}[tbp]
\includegraphics[bb=54 130 560 555, width=8 cm, clip]{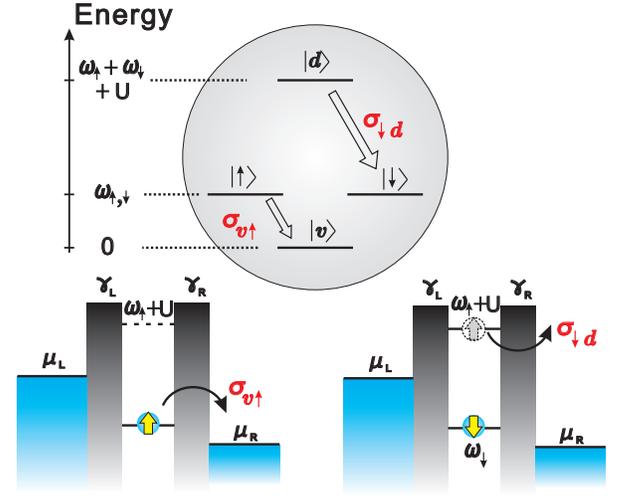}
\caption{{}(color online) Schematic illustration of Coulomb blockade
transport. The quantum dot can be regarded as a four-level
artificial atom with $|v\rangle, |\uparrow\rangle ,
|\downarrow\rangle $ and $|d\rangle$ representing the vacuum,
spin-up, spin-down, and the doubly occupied states. The lower panel
shows the two element processes of a spin-up electron tunneling out
of the quantum dot depending on whether the spin-down state is
occupied.} \label{FigScheme2}
\end{figure}

As has been discussed in Sec.\ref{Sec_projection}, though there is
interaction between the electrons in the system Hamiltonian,
$H_{\text{sys}}$ is diagonalized as
\begin{equation}
H_{\text{sys}}=\hbar \omega _{\uparrow
}\sigma_{\uparrow\uparrow}+\hbar \omega _{\downarrow
}\sigma_{\downarrow\downarrow}+\left(\hbar \omega _{\uparrow }+\hbar
\omega _{\downarrow }+U\right)\sigma_{dd},
\end{equation}%
and the projection operators are related to the Fermion operators by
\begin{subequations}
\label{CBFermiToProj}
\begin{eqnarray}
a_{\uparrow } &=&\sigma _{v\uparrow }-\sigma _{\downarrow d}, \\
a_{\downarrow } &=&\sigma _{v\downarrow }+\sigma _{\uparrow d},
\end{eqnarray}
\end{subequations}
where $\sigma_{ij}=|i\rangle\langle j|$ for
$i,j=v,\uparrow,\downarrow \text{ and } d $. The subscripts
$v,\uparrow ,\downarrow $ and $d$ represent the vacuum state,
spin-up, spin-down, and doubly occupied state, respectively
(Fig.~\ref{FigScheme2}). Annihilating an electron with definite spin
(say spin-up) from the quantum dot consists of two different
projection processes depending on whether the spin-down level is
occupied or not.

Here, we assume that the quantum dot is coupled to ferromagnetic
leads. Thus, the electron with different spin can tunnel on and off
the quantum dot with different rates. The quantum Langevin equations
of the projection operators $\sigma _{ij}$ in this Coulomb blockade
case follow the general formula in Sec.~\ref{Sec_projection}. The
resultant equations for the diagonal elements are%
\begin{widetext}
\begin{subequations}\label{LangevinEqDiag}
\begin{eqnarray}
\dot{\sigma}_{vv} &=&-\frac{\Gamma _{\uparrow }+\Gamma _{\downarrow }}{2}%
{\sigma}_{vv}+\frac{\Gamma _{\uparrow }}{2}{\sigma}_{\uparrow
\uparrow }+\frac{\Gamma _{\downarrow }}{2}{\sigma}_{\downarrow
\downarrow }-\left[ {\mathcal{F}}_{\uparrow }^{\dag }
{\sigma}_{v\uparrow }+{\mathcal{F}}_{\downarrow }^{\dag } {\sigma}_{v\downarrow }-{\mathcal{F}}%
_{\uparrow } {\sigma}_{\uparrow v}-%
{\mathcal{F}}_{\downarrow } {\sigma}%
_{\downarrow v}\right] ,  \label{LangevinEqOffdiag:a} \\
\dot{{\sigma}}_{\uparrow \uparrow } &=&-\frac{\Gamma _{\uparrow
}+\Gamma _{\downarrow }}{2}{\sigma}_{\uparrow \uparrow
}+\frac{\Gamma
_{\uparrow }}{2}{\sigma}_{vv}+\frac{\Gamma _{\downarrow }}{2}{%
\sigma}_{dd}+\left[ {\mathcal{F}}_{\uparrow }^{\dag }
{\sigma}_{v\uparrow }-{\mathcal{F}}_{\downarrow }^{\dag } {\sigma}_{\uparrow d}-{\mathcal{F}}%
_{\uparrow } {\sigma}_{\uparrow v}+%
{\mathcal{F}}_{\downarrow } {\sigma}%
_{d\uparrow }\right] ,  \label{LangevinEqOffdiag:b} \\
\dot{{\sigma}}_{\downarrow \downarrow } &=&-\frac{\Gamma _{\uparrow
}+\Gamma _{\downarrow }}{2}{\sigma}_{\downarrow \downarrow
}+\frac{\Gamma
_{\downarrow }}{2}{\sigma}_{vv}+\frac{\Gamma _{\uparrow }}{2}{%
\sigma}_{dd}+\left[ {\mathcal{F}}_{\uparrow }^{\dag }
{\sigma}_{\downarrow d}+{\mathcal{F}}_{\downarrow }^{\dag
} {\sigma}_{v\downarrow }-\mathcal{F%
}_{\uparrow } {\sigma}_{d\downarrow
}-{\mathcal{F}}_{\downarrow } {\sigma}%
_{\downarrow v}\right] , \\
\dot{{\sigma}}_{dd} &=&-\frac{\Gamma _{\uparrow }+\Gamma _{\downarrow }}{2}%
{\sigma}_{dd}+\frac{\Gamma _{\downarrow }}{2}{\sigma}_{\uparrow
\uparrow }+\frac{\Gamma _{\uparrow }}{2}{\sigma}_{\downarrow
\downarrow }-\left[ {\mathcal{F}}_{\uparrow }^{\dag }
{\sigma}_{\downarrow d}-{\mathcal{F}}_{\downarrow }^{\dag
} {\sigma}_{\uparrow d}-\mathcal{F%
}_{\uparrow } {\sigma}_{d\downarrow
}+{\mathcal{F}}_{\downarrow } {%
\sigma}_{d\uparrow }\right] ,
\end{eqnarray}%
\end{subequations}
and those for the off-diagonal elements are
\begin{subequations}
\label{LangevinEqOffdiag}
\begin{eqnarray}
\dot{{\sigma}}_{v\uparrow } &=&-i \Delta_{v\uparrow}\sigma_{v\uparrow }-\frac{\Gamma _{\uparrow }+\Gamma _{\downarrow }%
}{2}{\sigma}_{v\uparrow }-\frac{\Gamma _{\downarrow }}{2}{\sigma}%
_{\downarrow d}+\left[ {\mathcal{F}}_{\uparrow }\left( {\sigma}_{vv}+%
{\sigma}_{\uparrow \uparrow }\right) +{\mathcal{F}}_{\downarrow }{%
\sigma}_{\downarrow \uparrow }+{\mathcal{F}}_{\downarrow }^{\dag }{\sigma%
}_{vd}\right] , \\
\dot{{\sigma}}_{\downarrow d} &=&-i\Delta_{\downarrow
d}\sigma_{\downarrow d}-\frac{\Gamma _{\uparrow }+\Gamma
_{\downarrow }}{2}{\sigma}_{\downarrow d}-\frac{\Gamma _{\downarrow }}{%
2}{\sigma}_{v\uparrow }-\left[ {\mathcal{F}}_{\uparrow }\left( {%
\sigma}_{\downarrow \downarrow }+{\sigma}_{dd}\right) -{\mathcal{F}}%
_{\downarrow }{\sigma}_{\downarrow \uparrow
}-{\mathcal{F}}_{\downarrow
}^{\dag }{\sigma}_{vd}\right] , \\
\dot{{\sigma}}_{v\downarrow }
&=&-i\Delta_{v\downarrow}\sigma_{v\downarrow }-\frac{\Gamma
_{\uparrow }+\Gamma
_{\downarrow }}{2}{\sigma}_{v\downarrow }-\frac{\Gamma _{\uparrow }}{2}%
{\sigma}_{\uparrow d}+\left[ {\mathcal{F}}_{\downarrow }\left( {%
\sigma}_{vv}+{\sigma}_{\downarrow \downarrow }\right) +{\mathcal{F}}%
_{\uparrow }{\sigma}_{\uparrow \downarrow }-{\mathcal{F}}_{\uparrow
}^{\dag }{\sigma}_{vd}\right] , \\
\dot{{\sigma}}_{\uparrow d} &=&-i\Delta_{\uparrow d}\sigma_{\uparrow
d}-\frac{\Gamma _{\uparrow }+\Gamma _{\downarrow }%
}{2}{\sigma}_{\uparrow d}+\frac{\Gamma _{\uparrow }}{2}{\sigma}%
_{v\downarrow }+\left[ {\mathcal{F}}_{\downarrow }\left( {\sigma}%
_{\uparrow \uparrow }+{\sigma}_{dd}\right) -{\mathcal{F}}_{\uparrow }%
{\sigma}_{\uparrow \downarrow }+{\mathcal{F}}_{\uparrow }^{\dag }{%
\sigma}_{vd}\right] , \\
\dot{{\sigma}}_{vd} &=&-i\Delta_{vd}\sigma_{vd}
-\frac{\Gamma _{\uparrow }+\Gamma _{\downarrow }}{2}%
{\sigma}_{vd}+\left[ {\mathcal{F}}_{\uparrow }\left( {\sigma}%
_{\uparrow d}+{\sigma}_{v\downarrow }\right)
-{\mathcal{F}}_{\downarrow }\left(\sigma _{v\uparrow
}-{\sigma}_{\downarrow
d}\right) \right] , \\
\dot{{\sigma}}_{\uparrow \downarrow } &=&-i\Delta_{\uparrow
\downarrow}\sigma_{\uparrow \downarrow}-\frac{\Gamma _{\uparrow
}+\Gamma
_{\downarrow }}{2}{\sigma}_{\uparrow \downarrow }+\left[ {\mathcal{F}}%
_{\uparrow }^{\dag }\left( {\sigma}_{\uparrow d}+{\sigma}%
_{v\downarrow }\right) -{\mathcal{F}}_{\downarrow }\left( {\sigma}%
_{\uparrow v}-{\sigma}_{d\downarrow }\right) \right] ,
\end{eqnarray}%
\end{subequations}
\end{widetext}
where $\Delta_{i j}=\omega_j-\omega_i$,
$\mathcal{F}_s\equiv\mathcal{F}_s(t)=\mathcal{L}_s(t)+\mathcal{R}_s(t)$
for $s\in \{\uparrow ,\downarrow \} $, and the spin dependent noise
operators $\mathcal{L}_{s}(t)$ and $\mathcal{R}_{s}(t)$ are defined
as
\begin{subequations}
\begin{eqnarray}
\mathcal{L}_{s}(t) &=&-\sum_{k}\xi _{ks}e^{-i\omega
_{ks}^{(L)}t}b_{ks}\left(
0\right) , \\
\mathcal{R}_{s}(t) &=&-\sum_{j}\zeta _{js}e^{-i\omega
_{js}^{(R)}t}c_{js}\left( 0\right).
\end{eqnarray}%
The damping rate $\Gamma_s =\gamma _{Ls}+\gamma _{Rs}$, with the
spin dependent tunneling rates
\end{subequations}
\begin{subequations}
\begin{eqnarray}
\gamma_{Ls} &=&2\pi \xi_{s}^{2}\left( \omega
_{s}\right)D_{s}^{(L)}\left(
\omega _{s}\right), \\
\gamma_{Rs} &=&2\pi \zeta_{s}^{2}\left( \omega
_{s}\right)D_{s}^{(R)}\left( \omega _{s}\right) ,
\end{eqnarray}%
where $\xi_s$ and $\zeta_s$ are the coupling amplitudes of the
quantum dot to the left and right leads, and $D_{s}^{(L)}(\omega)$
and $D_{s}^{(R)}(\omega)$ are the spin-resolved density of states of
left and right leads, respectively.

These Langevin equations of the system variables are analogous to
the ones used to describe the quantum theory of
Laser\cite{bookHaken,bookScully}. In the quantum theory of Laser,
the atoms are subject to bosonic reservoirs, while in our quantum
transport case, the quantum dot is ``pumped'' by a fermionic
reservoir (the left lead), and output to another fermionic reservoir
(the right lead).

In contrast to the non-interacting case [see Eq.~(\ref{singellevelLagevinEq}%
)], the noise operators couple to the system projection operators in
 Eqs.~(\ref{LangevinEqDiag}) and (\ref%
{LangevinEqOffdiag}). The correlations between noise operators and
projection operators, such as $\langle\mathcal{L}_{s}^{\dag }(t)%
\sigma_{ij}(t)\rangle$, are calculated according to the generalized
causality relation Eq.~(\ref{causalitiesPOp}). Taking $\langle
\mathcal{L}_{\uparrow}^{\dag }(t)\sigma_{v\uparrow}(t)\rangle$ for
example,

\end{subequations}
\begin{eqnarray}  \label{causalityAver}
&&\langle \mathcal{L}_{\uparrow}^{\dag
}(t)\sigma_{v\uparrow}(t)\rangle=\langle\tilde{\mathcal{L}}_{\uparrow
}^{\dag }(t) \tilde{\sigma}
_{v\uparrow }(t)\rangle  \notag \\
&=&\langle[\tilde{\mathcal{L}}_{\uparrow }^{\dag }(t),
\tilde{\sigma}
_{v\uparrow }(t)]_{+}\rangle-\langle \tilde{\sigma} _{v\uparrow }(t)\tilde{%
\mathcal{L}}_{\uparrow }^{\dag }(t)\rangle  \notag \\
&=&\frac{1}{2}\gamma_{L}\langle[a_{\uparrow }^{\dag }(t), \sigma
_{v\uparrow }(t)]_{+}\rangle
=\frac{1}{2}\gamma_{L}\left\langle\sigma _{vv }(t)+\sigma
_{\uparrow\uparrow }(t)\right\rangle,
\end{eqnarray}
where $\tilde{\sigma} _{v\uparrow }(t)=\sigma _{v\uparrow
}(t)e^{i\omega_{\uparrow}t}$ is the slow-varying amplitude of
projection operator, and $\tilde{\mathcal{L}}_{\uparrow }^{\dag
}(t)={\mathcal{L}}_{\uparrow
}^{\dag }(t)e^{-i\omega_{\uparrow}t}$. The correlation $\langle \tilde{\sigma} _{v\uparrow }(t)\tilde{\mathcal{L}}%
_{\uparrow }^{\dag }(t)\rangle$ in the second line of Eq.~(\ref%
{causalityAver}) vanishes, when the noise operator $\tilde{\mathcal{L}}%
_{\uparrow }^{\dag }(t)$ acts on the full-filled fermi sea of the
left lead, see Appendix \ref{appendix_A}.

With these correlations, ensemble average of
Eqs.~(\ref{LangevinEqDiag}) and (\ref{LangevinEqOffdiag}) gives the
\textquotedblleft rate equations\textquotedblright\ for the diagonal
elements:
\begin{subequations}
\label{rateEq}
\begin{eqnarray}
\langle \dot{\sigma}_{vv}\rangle &=&-\left( \gamma _{L\uparrow
}+\gamma _{L\downarrow }\right) \langle \sigma _{vv}\rangle +\gamma
_{R\uparrow }\langle \sigma _{\uparrow \uparrow }\rangle +\gamma
_{R\downarrow }\langle
\sigma _{\downarrow \downarrow }\rangle , \\
\langle \dot{\sigma}_{\uparrow \uparrow }\rangle &=&-\gamma
_{R\uparrow }\langle \sigma _{\uparrow \uparrow }\rangle +\gamma
_{L\uparrow }\langle \sigma _{vv}\rangle +\left( \gamma
_{L\downarrow }+\gamma _{R\downarrow
}\right) \langle \sigma _{dd}\rangle , \\
\langle \dot{\sigma}_{\downarrow \downarrow }\rangle &=&-\gamma
_{R\downarrow }\langle \sigma _{\downarrow \downarrow }\rangle
+\gamma _{L\downarrow }\langle \sigma _{vv}\rangle +\left( \gamma
_{L\uparrow
}+\gamma _{R\uparrow }\right) \langle \sigma _{dd}\rangle , \\
\langle \dot{\sigma}_{dd}\rangle &=&-\left( \gamma _{L\uparrow
}+\gamma _{L\downarrow }+\gamma _{R\uparrow }+\gamma _{R\downarrow
}\right) \langle \sigma _{dd}\rangle ,
\end{eqnarray}%
and for off-diagonal elements:
\end{subequations}
\begin{subequations}
\label{offdiagAver}
\begin{eqnarray}
\left\langle \dot{\sigma}_{v\uparrow }\right\rangle &=&-\left(
i\Delta _{v\uparrow }+\Gamma _{1}\right) \left\langle \sigma
_{v\uparrow
}\right\rangle -\frac{\gamma _{L\downarrow }+2\gamma _{R\downarrow }}{2}%
\left\langle \sigma _{\downarrow d}\right\rangle , \\
\left\langle \dot{\sigma}_{\downarrow d}\right\rangle &=&-\left(
i\Delta _{\downarrow d}+\Gamma _{2}\right) \left\langle \sigma
_{\downarrow d}\right\rangle -\frac{\gamma _{L\downarrow
}}{2}\left\langle \sigma
_{v\uparrow }\right\rangle , \\
\left\langle \dot{\sigma}_{v\downarrow }\right\rangle &=&-\left(
i\Delta _{v\downarrow }+\Gamma _{3}\right) \left\langle \sigma
_{v\downarrow
}\right\rangle +\frac{\gamma _{L\uparrow }+2\gamma _{R\uparrow }}{2}%
\left\langle \sigma _{\uparrow d}\right\rangle , \\
\left\langle \dot{\sigma}_{\uparrow d}\right\rangle &=&-\left(
i\Delta _{\uparrow d}+\Gamma _{4}\right) \left\langle \sigma
_{\uparrow d}\right\rangle +\frac{\gamma _{L\uparrow
}}{2}\left\langle \sigma
_{v\downarrow }\right\rangle , \\
\left\langle \dot{\sigma}_{vd}\right\rangle &=&-\left( i\Delta
_{vd}+2\gamma _{L\uparrow }+2\gamma _{L\downarrow }+\gamma
_{R\uparrow }+\gamma
_{R\downarrow }\right) \left\langle \sigma _{vd}\right\rangle , \\
\left\langle \dot{\sigma}_{\uparrow \downarrow }\right\rangle
&=&-\left( i\Delta _{\uparrow \downarrow }+\gamma _{R\uparrow
}+\gamma _{R\downarrow }\right) \left\langle \sigma _{\uparrow
\downarrow }\right\rangle ,
\end{eqnarray}%
where $\Gamma _{1}=(\gamma _{L\uparrow }+\gamma _{L\downarrow
}+\gamma _{R\uparrow })/2$, $\Gamma _{2}=(\gamma _{L\uparrow
}+\gamma _{L\downarrow }+\gamma _{R\uparrow }+2\gamma _{R\downarrow
})/2$, $\Gamma _{3}=(\gamma _{L\uparrow }+\gamma _{L\downarrow
}+\gamma _{R\downarrow })/2$ and $\Gamma _{4}=(\gamma _{L\uparrow
}+\gamma _{L\downarrow }+2\gamma _{R\uparrow }+\gamma _{R\downarrow
})/2$. Eq.~(\ref{offdiagAver}) shows that the coherence between
energy levels vanish after a long time, i.e.
\end{subequations}
\begin{equation}
\langle \sigma _{ij}(t)\rangle =0,\text{ for }i\neq j\text{, and }%
t\rightarrow +\infty .  \label{SCvaloff}
\end{equation}%
This indicates the two different spin channels are incoherent, which
physically arises from the fact that noise operators with different
spins
are uncorrelated, i.e. $\langle \mathcal{L}_{\uparrow }^{\dag }(t)%
\mathcal{L}_{\downarrow }(t^{\prime })\rangle =0$.

The rate equations (\ref{rateEq}) describe the population transfer
between each energy levels, and the steady state populations are
\begin{subequations}
\label{SCvaldiag}
\begin{eqnarray}
\langle \sigma _{vv}\rangle  &=&\frac{\gamma _{R\uparrow }\gamma
_{R\downarrow }}{\gamma _{L\downarrow }\gamma _{R\uparrow }+\gamma
_{L\uparrow }\gamma _{R\downarrow }+\gamma _{R\uparrow }\gamma
_{R\downarrow
}}, \\
\langle \sigma _{\uparrow \uparrow }\rangle  &=&\frac{\gamma
_{L\uparrow }\gamma _{R\downarrow }}{\gamma _{L\downarrow }\gamma
_{R\uparrow }+\gamma _{L\uparrow }\gamma _{R\downarrow }+\gamma
_{R\uparrow }\gamma _{R\downarrow
}}, \\
\langle \sigma _{\downarrow \downarrow }\rangle  &=&\frac{\gamma
_{L\downarrow }\gamma _{R\uparrow }}{\gamma _{L\downarrow }\gamma
_{R\uparrow }+\gamma _{L\uparrow }\gamma _{R\downarrow }+\gamma
_{R\uparrow
}\gamma _{R\downarrow }}, \\
\langle \sigma _{dd}\rangle  &=&0.
\end{eqnarray}%
The average current is
\end{subequations}
\begin{eqnarray}\label{AverCurrentCB}
&&\langle \hat{I}_{R}\rangle =\langle \hat{I}_{R\uparrow }\rangle
+\langle
\hat{I}_{R\downarrow }\rangle   \notag  \label{CBcurrent} \\
&=&\gamma _{R\uparrow }\langle \sigma _{\uparrow \uparrow }\rangle
+\gamma
_{R\downarrow }\langle \sigma _{\downarrow \downarrow }\rangle   \notag \\
&=&\frac{\left( \gamma _{L\uparrow }+\gamma _{L\downarrow }\right)
\gamma _{R\uparrow }\gamma _{R\downarrow }}{\gamma _{L\downarrow
}\gamma _{R\uparrow }+\gamma _{L\uparrow }\gamma _{R\downarrow
}+\gamma _{R\uparrow }\gamma _{R\downarrow }}.
\end{eqnarray}%
The current vanishes if $\gamma_{R\uparrow}=0$ or
$\gamma_{R\downarrow}=0$. This is because turning off a certain spin
channel, say the spin-up channel, i.e. $\gamma_{R\uparrow}=0$, will
induce the accumulation of the spin-up electron on the quantum dot.
Then the electron tunneling of both spin channels is blockade due to
the strong Coulomb interaction. When the tunneling rates are
spin-independent, i.e. $\gamma _{L\uparrow }=\gamma _{L\downarrow
}=\gamma _{L}$ and $\gamma _{R\uparrow }=\gamma _{R\downarrow
}=\gamma _{R}$, the average current in Eq.~(\ref{AverCurrentCB})
becomes $\langle \hat{I}_{R}\rangle =2\gamma _{L}\gamma
_{R}/(2\gamma _{L}+\gamma _{R})$, which accords with the results
obtained by other
methods.\cite{GurvitzPRBMasterEq,LiXinQiCurrent,LiXinQiNoise}

\subsection{Current noise}

Now we turn to the current noise. Similar to the non-interacting
case, the two-time correlations of the following form should be
calculated
\begin{subequations}
\label{GECorrelationCB}
\begin{eqnarray}
&& \langle n_{s}\left( t\right) n_{s'}\left( t+\tau \right)\rangle,
\label{GECorrelationCB1} \\
&& \langle a_{s}^{\dag }\left( t\right) \mathcal{R}_{s}\left( t\right)%
\mathcal{R}_{s'}^{\dag }\left( t+\tau \right) a_{s'}\left( t+\tau
\right)\rangle, \\
&& \langle a_{s}^{\dag }\left( t\right) \mathcal{R}_{s}\left(
t\right) a_{s'}^{\dag}\left( t+\tau \right) a_{s'}\left(
t+\tau\right)\rangle.
\end{eqnarray}
In the interacting case, the system projection operators cannot be
expressed in terms of the simple integration of the noise operators
as in the non-interacting case, see Eq.~(\ref{SinglelevelSolution}).
The causality relations introduced in Sec \ref{Sec:causality}
provide us a convenient way to convert the noise-system correlation
to the system-system correlation. Thus, in the white-noise limit, as
a powerful tool, the quantum regression theorem is applied to
calculate the two-time system correlations.

Noticing that the noise operator ${\mathcal{R}}_{s}(t)$ plays the
role of ``annihilation operator'', the correlations between noise
and system operators can be calculated following the spirit of the
Wick's theorem(see Appendix~\ref{appendix_A}). Taking the spin-up
component for example, the correlation $\langle a_{\uparrow }^{\dag
}\left( t\right) \mathcal{R}_{\uparrow }\left( t\right)
\mathcal{R}_{\uparrow }^{\dag }\left( t+\tau \right) a_{\uparrow
}\left( t+\tau \right) \rangle$ is
\end{subequations}
\begin{eqnarray}  \label{CB_Corr1}
&&\langle a_{\uparrow }^{\dag }\left( t\right) \mathcal{R}_{\uparrow
}\left( t\right) \mathcal{R}_{\uparrow }^{\dag }\left( t+\tau
\right) a_{\uparrow
}\left( t+\tau \right) \rangle  \notag \\
&=&\langle a_{\uparrow }^{\dag }\left(
t\right)[\mathcal{R}_{\uparrow }\left(
t\right),\mathcal{R}_{\uparrow }^{\dag }\left( t+\tau \right)]_{+}
a_{\uparrow }\left( t+\tau \right) \rangle  \notag \\
&&-\langle a_{\uparrow }^{\dag }\left( t\right)
\mathcal{R}_{\uparrow }^{\dag }\left( t+\tau \right)
\mathcal{R}_{\uparrow }\left( t\right)
a_{\uparrow }\left( t+\tau \right) \rangle  \notag \\
&=&\gamma_R\langle a_{\uparrow }^{\dag }\left( t\right) a_{\uparrow
}\left( t+\tau \right) \rangle\delta(\tau).
\end{eqnarray}
The second line of Eq.~(\ref{CB_Corr1}) is simplified by noticing the fact $[%
\mathcal{R}_{\uparrow }\left( t\right),\mathcal{R}_{\uparrow }^{\dag
}\left( t^{\prime }\right)]_{+}=\gamma_R\delta(t-t^{\prime })$, and
the third line vanishes since $[a_{\uparrow }^{\dag }\left(
t\right),\mathcal{R}_{\uparrow }^{\dag }\left( t+\tau
\right)]_{+}=0$. This white-noise correlation provides a constant
current noise background. Due to the $\delta(\tau)$ function, only
equal-time correlation ($\tau=0$) is relevant. By noticing
Eq.~(\ref{CBcurrent}), this correlation is written as
\begin{equation}
\langle a_{\uparrow }^{\dag }\left( t\right) \mathcal{R}_{\uparrow
}\left( t\right) \mathcal{R}_{\uparrow }^{\dag }\left( t+\tau
\right) a_{\uparrow }\left( t+\tau \right)\rangle=\langle
\hat{I}_{R\uparrow}\rangle\delta(\tau).
\end{equation}

For the correlation $\langle a_{\uparrow }^{\dag }\left( t\right) \mathcal{R}%
_{\uparrow }\left( t\right) a_{\uparrow }^{\dag }\left( t+\tau
\right) a_{\uparrow }\left( t+\tau \right) \rangle\equiv\langle
a_{\uparrow }^{\dag }\left( t\right) \mathcal{R}_{\uparrow }\left(
t\right) n_{\uparrow }\left( t+\tau \right) \rangle$, it can be
translated into the correlations between the system operators using
the causality relations
\begin{eqnarray}
&&\langle a_{\uparrow }^{\dag }\left( t\right) \mathcal{R}_{\uparrow
}\left(
t\right) n_{\uparrow }\left( t+\tau \right) \rangle  \notag \\
&=&\langle a_{\uparrow }^{\dag }\left( t\right)
[\mathcal{R}_{\uparrow
}\left( t\right), n_{\uparrow }\left( t+\tau \right)] \rangle  \notag \\
&=&\gamma_R\langle a_{\uparrow }^{\dag }\left( t\right)
[a_{\uparrow}\left(
t\right), n_{\uparrow }\left( t+\tau \right)] \rangle  \notag \\
&=&\gamma_R\langle n_{\uparrow}\left( t\right)n_{\uparrow }\left(
t+\tau \right) \rangle-\gamma_R\langle a_{\uparrow }^{\dag }\left(
t\right)n_{\uparrow }\left( t+\tau \right) a_{\uparrow}\left(
t\right) \rangle.
\end{eqnarray}%
The first term cancels out the contribution of
Eq.~(\ref{GECorrelationCB1}). As a result, the spin-up
current-current correlation is
\begin{eqnarray}  \label{CBcurrentCorr}
&&g_{\uparrow \uparrow}^{\left( 2\right) }\left( \tau \right)
=\lim_{t\rightarrow +\infty }\langle \hat{I}_{R\uparrow }\left(
t\right) \hat{I}_{R\uparrow }\left( t+\tau \right) \rangle\notag\\
&=&\lim_{t\rightarrow +\infty }\langle
\hat{I}_{R\uparrow}\rangle\delta(\tau)+\gamma_{R\uparrow}^{2}\langle
a_{\uparrow }^{\dag }\left( t\right)n_{\uparrow }\left( t+\tau
\right) a_{\uparrow}\left( t\right) \rangle.
\end{eqnarray}
The current-current correlations of different spin components are
calculated similarly, and in general, they can expressed in terms of
the correlations of system operators as
\begin{eqnarray}  \label{CBcurrentCorrGeneral}
&&g_{s s'}^{\left( 2\right) }\left( \tau \right) =\lim_{t\rightarrow
+\infty }\langle \hat{I}_{R s }\left(
t\right) \hat{I}_{R s'}\left( t+\tau \right) \rangle\notag\\
&=&\lim_{t\rightarrow +\infty }\langle \hat{I}_{R
s}\rangle\delta(\tau)\delta_{s s'}+\gamma_{R s}\gamma_{R s'}\langle
a_{s}^{\dag }\left( t\right)n_{s'}\left( t+\tau \right) a_{s}\left(
t\right) \rangle.
\end{eqnarray}
Here we have shown an analogous form of the current-current
correlation to the second order optical coherence
function.\cite{bookScully} The last term of
Eq.~(\ref{CBcurrentCorrGeneral}) can be calculated from the quantum
regression theorem. By this theorem, the current-current correlation
function is determined by the rate equations (\ref{rateEq}), and in
the Coulomb blockade case, it does not show the effect of the
quantum coherence terms in Eq.~(\ref{offdiagAver}).

The total current correlation function is
\begin{eqnarray}
&&g^{\left( 2\right) }\left( \tau \right) = \lim_{t\rightarrow
+\infty }\langle \hat{I}_{R}\left( t\right) \hat{I}_{R}\left( t+\tau
\right) \rangle
-\langle \hat{I}_{R}\rangle ^{2}  \notag \\
&=& g_{\uparrow\uparrow}^{\left( 2\right) }\left( \tau
\right)+g_{\uparrow\downarrow}^{\left( 2\right) }\left( \tau
\right)+g_{\downarrow\uparrow}^{\left( 2\right) }\left( \tau
\right)+g_{\downarrow\downarrow}^{\left( 2\right) }\left( \tau
\right)-\langle \hat{I}_{R}\rangle ^{2}
\end{eqnarray}
Its Fourier transformation gives the current noise spectrum
$S(\omega)$. In the spin independent tunneling rate case, i.e.
$\gamma _{L\uparrow }=\gamma _{L\downarrow }=\gamma _{L}$ and
$\gamma _{R\uparrow }=\gamma _{R\downarrow }=\gamma _{R}$, the noise
spectrum is
\begin{eqnarray}
S(\omega)=2e\langle
\hat{I}_{R}\rangle\frac{4\gamma_L^2+3\gamma_L\gamma_R+\gamma_R^2}{(2\gamma_L+\gamma_R)^2+\omega^2}
\end{eqnarray}
This result deviates from the single-level case [see
Eq.~(\ref{singleCurrentNoise})], due to the presence of the Coulomb
interaction.

\begin{figure}[tbp]
\includegraphics[bb=50 50 434 725, width=7 cm, clip]{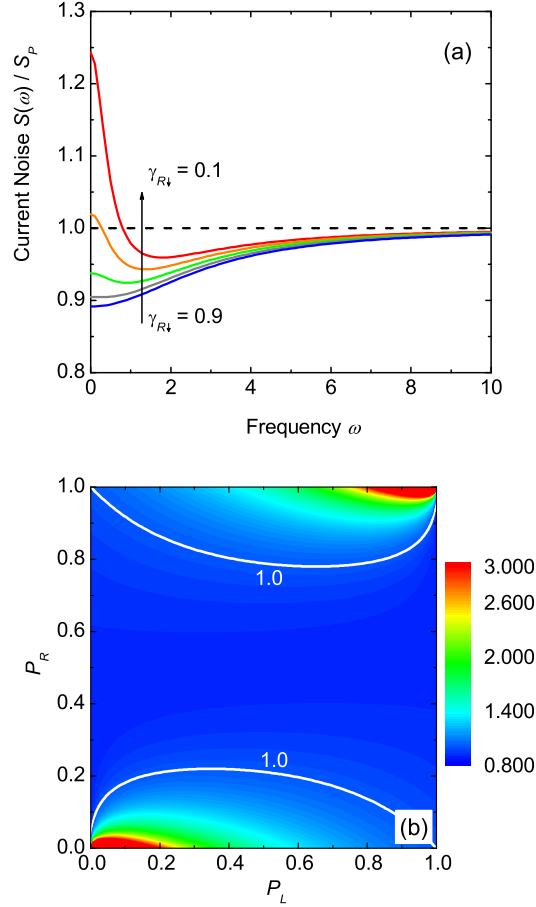}
\caption{(color online) (a) Current noise spectra $S(\omega)$
(normalized by the Poisson value $S_{P}=2e\langle
\hat{I}_{R}\rangle$) for $\gamma_{R\downarrow}=0.1, 0.3, \cdots,
0.9$. Other parameters are chosen as
$\gamma_{L\uparrow}=\gamma_{L\downarrow}=\gamma_{R\uparrow}=1$. (b)
Fano factor as a function of the imbalance between spin-resolved
tunneling rates $P_L$ and $P_R$, which are defined as
$P_L=\gamma_{L\uparrow}/(\gamma_{L\uparrow}+\gamma_{L\downarrow})$
and
$P_R=\gamma_{R\uparrow}/(\gamma_{R\uparrow}+\gamma_{R\downarrow})$,
for given total tunneling rates
$\gamma_{L\uparrow}+\gamma_{L\downarrow}=\gamma_{R\uparrow}+\gamma_{R\downarrow}=1$.
The white thick lines are the boundary between sub-Poisson and
super-Poisson regime, i.e. $F=1$.}\label{FigNoiseSpectra}
\end{figure}

Typical current noise spectra for the spin dependent tunneling rate
case are shown in Fig.~(\ref{FigNoiseSpectra}a). The Fano factor is
\begin{eqnarray}
&&F\equiv S(\omega=0)/2e\langle \hat{I}_R\rangle\notag\\
&=&1-\frac{\gamma _{R\uparrow }\gamma _{R\downarrow }(\gamma
_{L\downarrow }\gamma _{R\uparrow }+\gamma _{L\uparrow }\gamma
_{R\downarrow })-\gamma _{L\uparrow }\gamma _{L\downarrow }(\gamma
_{R\uparrow }-\gamma _{R\downarrow })^2}{2(\gamma _{L\downarrow
}\gamma _{R\uparrow }+\gamma _{L\uparrow }\gamma _{R\downarrow
}+\gamma _{R\uparrow }\gamma _{R\downarrow })^2}.
\end{eqnarray}
It is found that super-Poissonian noise arises when the tunneling is
spin dependent, which can be realized, e.g., by using magnetized
barriers between the leads and the quantum dot. Super-Poissonian
noise appears when the numerator of the second term becomes
negative. The Fano factor as a function of the tunneling rate
imbalance is shown in Fig.~(\ref{FigNoiseSpectra}b).

Physically, the super-Poissonian noise is the consequence of the
dynamical channel blockade effect.\cite{SuperPoissonBruder,FSC} The
tunneling rate imbalance induces different average currents for the
two spin channels. Thus, in additional to the noises of each
channels themselves, the shot noise between the two channels gives
rise to the low frequency noise enhancement. Such kind of shot noise
is absent when $P_L=1-P_R$ since the two spin channels have the same
current.
\section{Relation to Other Theories}

\label{comparison}

\subsection{Relation to Landauer-B\"{u}ttiker formula}

Here, we show that the Landauer-B\"{u}ttiker formula can be reproduced by the quantum Langevin
approach. For simplicity, let us consider the single energy level
transport example.

According to Eq.~(\ref{CurrentOPDef}) and the boundary relations
Eq.~(\ref{boundaryRelation1}), the current operators can be expressed
solely by the input and output noise operators. For example,
\begin{equation}
\hat{I}_{R}=\frac{1}{\gamma_{R}}\left(\tilde{\mathcal{R}}^{\dag}_{\text{out}%
}\left( t\right)\tilde{\mathcal{R}}_{\text{out}}\left( t\right)-\tilde{%
\mathcal{R}}^{\dag}_{\text{in}}\left( t\right)\tilde{\mathcal{R}}_{\text{in}%
}\left( t\right)\right).
\end{equation}
Thus, it is clear that the average current is divided into the input current
proportional to $\langle\tilde{\mathcal{R}}^{\dag}_{\text{in}}\left( t\right)%
\tilde{\mathcal{R}}_{\text{in}}\left( t\right)\rangle$ and the output
current proportional to $\langle\tilde{\mathcal{R}}^{\dag}_{\text{out}%
}\left( t\right)\tilde{\mathcal{R}}_{\text{out}}\left( t\right)\rangle$.

Furthermore, defining the scattering matrix $\mathbf{S}$, the
Fourier transformation of output noise operators is expressed in
terms of the input operators as
\begin{equation}
\left(
\begin{array}{c}
\tilde{\mathcal{L}}_{\text{out}}\left( \omega\right) \\
\tilde{\mathcal{R}}_{\text{out}}\left( \omega\right)%
\end{array}
\right)=\mathbf{S}\left(
\begin{array}{c}
\tilde{\mathcal{L}}_{\text{in}}\left( \omega\right) \\
\tilde{\mathcal{R}}_{\text{in}}\left( \omega\right)%
\end{array}
\right),  \label{TransMatrixRelation}
\end{equation}
with
\begin{subequations}
\begin{eqnarray}
\tilde{\mathcal{L}}_{\text{in/out}}\left(
\omega\right)&=&\int_{-\infty}^{\infty} e^{i \omega t}\tilde{\mathcal{L}}_{%
\text{in/out}}\left( t\right)dt, \\
\tilde{\mathcal{R}}_{\text{in/out}}\left( \omega\right) &=&
\int_{-\infty}^{\infty} e^{i \omega t}\tilde{\mathcal{R}}_{\text{in/out}%
}\left( t\right)dt,
\end{eqnarray}
\end{subequations}
and 
\begin{eqnarray}
\mathbf{S}(\omega)&\equiv& \left(
\begin{array}{cc}
\mathbb{R}_{L\leftarrow L}(\omega) & \mathbb{T}_{L\leftarrow R}(\omega) \\
\mathbb{T}_{R\leftarrow L}(\omega) & \mathbb{R}_{R\leftarrow R}(\omega)%
\end{array}
\right)\notag\\
&=& \frac{2}{\gamma_L+\gamma_R-2i\omega}\left(
\begin{array}{cc}
\frac{\gamma_L-\gamma_R}{2}-i \omega & \gamma_L \\
\gamma_R & -\frac{\gamma_L-\gamma_R}{2}+i \omega%
\end{array}
\right),
\end{eqnarray}
where the functions $\mathbb{T}_{i\leftarrow j}(\omega)$ and $\mathbb{R}%
_{i\leftarrow j}(\omega)$ can be regarded as the energy dependent
transmission and reflection coefficients from lead $j$ to lead $i$.
The Fourier transformation of the average current is
\begin{eqnarray}
&&\langle\hat{I}_{R}(\omega)\rangle=\int
\langle\hat{I}_{R}(t)\rangle e^{i
\omega t} dt \notag \\
&=&\frac{1}{\gamma_{R}} \int_{-\infty}^{\infty} \left[\langle%
\tilde{\mathcal{R}}^{\dag}_{\text{out}}\left( \omega^{\prime }\right)\tilde{%
\mathcal{R}}_{\text{out}}\left( \omega^{\prime
}+\omega\right)\rangle\right.\notag\\
&-&\left.\langle\tilde{\mathcal{R}}^{\dag}_{\text{in}}\left( \omega^{\prime }\right)%
\tilde{\mathcal{R}}_{\text{in}}\left( \omega^{\prime }+\omega\right)\rangle%
\right]\frac{d\omega^{\prime }}{{2\pi}}.
\end{eqnarray}
Noticing the relation Eq.~(\ref{TransMatrixRelation}) and the
correlations between the noise operators
\begin{subequations}
\begin{eqnarray}
\langle\tilde{\mathcal{L}}^{\dag}_{\text{in}}\left( \omega^{\prime }\right)%
\tilde{\mathcal{L}}_{\text{in}}\left( \omega^{\prime }+\omega\right)\rangle
&=& 2\pi\gamma_{L}\delta(\omega), \\
\langle\tilde{\mathcal{R}}^{\dag}_{\text{in}}\left( \omega^{\prime }\right)%
\tilde{\mathcal{R}}_{\text{in}}\left( \omega^{\prime }+\omega\right)\rangle
&=& 0,
\end{eqnarray}
\end{subequations}
we obtain the Landauer-B\"{u}ttiker-like formula of the average
current
\begin{equation}
\langle\hat{I}_{R}\rangle = \int_{-\infty}^{+\infty} T(\omega^{\prime })%
\frac{d\omega^{\prime }}{2\pi}=\frac{\gamma_L\gamma_R}{\gamma_L+\gamma_R}.
\end{equation}
with the transmission spectrum
\begin{equation}
T(\omega)=\frac{\gamma_L\gamma_R}{(\frac{\gamma_L+\gamma_R}{2})^2+\omega^2}.
\end{equation}

\subsection{Relation to non-equilibrium Green's function theory}

Here we discuss the relation between the quantum Langevin approach
and the non-equilibrium Green's function theory for the quantum
transport
problems. The retarded Green's function is defined as\cite%
{PRLgreenFunc}
\begin{equation}
G_{s}(\tau)=-i
\theta(\tau)\left\langle\{a_{s}(t+\tau),a_{s}^{\dag}(t)\}\right\rangle,
\label{GreenFunc}
\end{equation}
for $s=\uparrow$ or $\downarrow$, from which the local density of states (LDOS) $%
\mathcal{D}_s(\omega)$ is given by
\begin{equation}
\mathcal{D}_{s}(\omega)=-\frac{1}{\pi}\text{Im}[G_{s}(\omega)],
\end{equation}
where $\tilde{G}_{s}(\omega)$ is the Fourier transformation of
$G_{s}(\tau)$. The LDOS contains the essential information about the
system relevant in quantum transport.
In the following, we take the Coulomb blockade example, and give the
retarded Green's function and the LDOS using the quantum noise
approach.

\begin{figure}[tbp]
\includegraphics[bb=30 20 540 400, width=7 cm, clip]{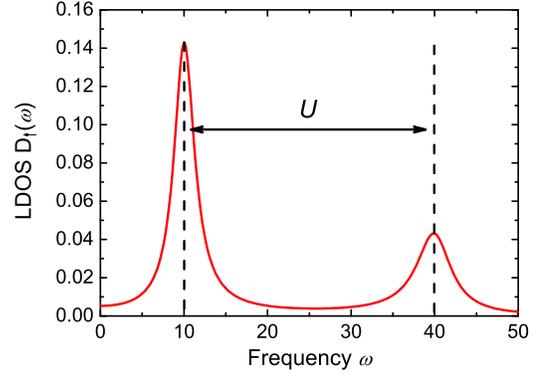}
\caption{{}(color online) Local density of states (LDOS) for the
Coulomb blockade case obtained by the quantum Langevin approach.
Parameters are $\omega_{\uparrow}=10$, $U=30$ and $\gamma=1$. LDOS shows a double-peak structure for $U\gg%
\protect\gamma$.} \label{FigDOS}
\end{figure}

Noticing that the definition of the retarded Green's function Eq.~(\ref%
{GreenFunc}) only involves the system operators $a_{s}(t)$ and $%
a_{s}^{\dag}(0)$, we apply the quantum regression theorem to
calculate their correlations. The retarded Green's function can be
expressed in terms of the two-time correlations between the
projection operators. Consider the spin-up component for example,
\begin{eqnarray}
G_{\uparrow}(\tau)&=&-i\theta(\tau)\left\langle \big\{\sigma _{v\uparrow
}(t+\tau)-\sigma _{\downarrow d}(t+\tau),\right.  \notag \\
&&\left.\sigma _{\uparrow v }(t)-\sigma _{d\downarrow }(t)\big\}%
\right\rangle.  \label{GreenFuncProj}
\end{eqnarray}
The equations of motion for these projection operators are given in
Eq.~(\ref{offdiagAver}). By the quantum regression theorem, the
two-time correlations are determined by
\begin{equation}
\frac{d}{d\tau} \left(
\begin{array}{c}
\langle\sigma_{\uparrow v}(t)\sigma_{v\uparrow }\left( t+\tau\right)\rangle
\\
\langle\sigma_{\uparrow v}(t)\sigma_{\downarrow d}\left( t+\tau\right)\rangle%
\end{array}%
\right) =\mathbf{M}\left(
\begin{array}{c}
\langle\sigma_{\uparrow v}(t)\sigma_{v\uparrow }\left( t+\tau\right)\rangle
\\
\langle\sigma_{\uparrow v}(t)\sigma_{\downarrow d}\left( t+\tau\right)\rangle%
\end{array}%
\right),  \label{CBFastQuantumLangevinAverRegress}
\end{equation}
with the initial condition for $\tau=0$
\begin{equation}
\left(
\begin{array}{c}
\langle\sigma_{\uparrow v}(t)\sigma_{v\uparrow }\left( t\right)\rangle \\
\langle\sigma_{\uparrow v}(t)\sigma_{\downarrow d}\left( t\right)\rangle%
\end{array}%
\right) = \left(
\begin{array}{c}
\langle\sigma_{\uparrow \uparrow}(t)\rangle \\
0%
\end{array}%
\right),
\end{equation}
where the coefficient matrix $\mathbf{M}$ is defined as
\begin{equation}
\mathbf{M}=\left(%
\begin{array}{cc}
-3\gamma/2-i\omega_{\uparrow} & -3\gamma/2 \\
-\gamma/2 & -5\gamma/2-i(\omega_{\uparrow}+U)%
\end{array}%
\right).  \label{matM}
\end{equation}
Here,
$\gamma_{L\uparrow}=\gamma_{L\downarrow}=\gamma_{R\uparrow}=\gamma_{R\downarrow}=\gamma$
is assumed for simplicity. The other correlations involved in
Eq.~(\ref{GreenFuncProj}) can be similarly calculated.

Thus the retarded Green's function is
\begin{eqnarray}
G_{\uparrow}(\tau)=-i\theta(\tau)e^{-2\gamma \tau}\left(W_{+}e^{-
i\omega_{+}\tau }+W_{-}e^{- i\omega_{-}\tau }\right),
\end{eqnarray}
with the renormalized frequencies
\begin{equation}
\omega_{\pm}=\omega_{\uparrow}+\frac{U}{2}\pm\frac{1}{2}\sqrt{U^2-2i
\gamma U-4\gamma^2},
\end{equation}
and the weight factors
\begin{equation}
W_{\pm}=\frac{1}{2}\pm\frac{U/6-i\gamma}{2\sqrt{U^2-2i\gamma
U-4\gamma^2}}.
\end{equation}
The Fourier transformation of the Green's function gives the LDOS (see Fig.~%
\ref{FigDOS}). It is obvious that, for the large $U$ case considered
in this paper, the LDOS consists of two Lorentz shape peaks,
centered around $\omega_{\uparrow}$ and $\omega_{\uparrow}+U$. The
two peaks separate from each other by $U$, which is a signature of
the Coulomb blockade.\cite{bookDatta2}

\section{Conclusions and Outlooks}

\label{conclusion}

In this paper, we have developed a quantum noise approach to treat
the quantum transport through a nanostructure such as a quantum dot.
We formulate the average current and the current noise in terms of
the correlations between the noise operators. The quantum noise
approach is applied to a paradigmatic example, namely, transport
through a single quantum dot under large biases and both the
non-interacting and Coulomb blockade cases are investigated. With the Markovian
approximation for the tunneling processes, the electron-electron interaction
in the quantum dot can be exactly treated.

The quantum noise approach provides a bridge between quantum optics
and quantum transport. Thus notions and methods in the quantum
optics could be adopted to study quantum transport through
nanostructures. Although we show the application of the quantum
noise approach by a single quantum dot example, the theory is not
limited to this simple case. On one hand, the system could be
generalized to more complicated ones, such as coupled quantum dots,
multi-end nano-circuits, or systems with spin interaction. On the
other hand, the reservoirs of other kinds, such as phonon baths or
spin baths, could be included to explore how such reservoirs would
affect the current and current noise, providing a method of studying
the bath dynamics via current noises. The Markovian approximation
may also be released with colored noise correlation functions of the
reservoir used in lieu of the white-noise model adopted in this
paper.

\begin{appendix}
\section{Properties of the Noise Operators}
\label{appendix_A}
In this appendix, we give the correlations
between noise operators.
We consider the single-level case here. The physical quantities of
interest are determined by the noise correlations such as $\langle
\tilde{\mathcal{L}}^{\dag }\left( t\right) \tilde{\mathcal{L}}\left(
t^{\prime }\right) \rangle$. According to the definition of the
noise operators,
\begin{eqnarray}
 && \langle \tilde{\mathcal{L}}^{\dag }\left( t\right)
\tilde{\mathcal{L}}\left( t^{\prime }\right) \rangle  \notag \\
&=&\sum_{k,k^{\prime }}\xi _{k}\xi _{k^{\prime }}e^{i(\omega
_{k}-\omega_0)t-i(\omega _{k^{\prime }}-\omega_0)t^{\prime }}\langle
b_{k}^{\dag
}b_{k^{\prime }}\rangle \notag \\
& =& \sum_{k}\xi _{k}^{2}e^{i(\omega _{k}-\omega_0)( t-t^{\prime })
}n
_{\operatorname{th}}^{(L)}\left( \omega _{k}\right)\notag \\
&=&\int_{0}^{\infty}\xi^{2}(\omega_k)D({\omega_k})n
_{\operatorname{th}}^{(L)}\left( \omega _{k}\right)e^{i(\omega
_{k}-\omega_0)( t-t^{\prime }) }d\omega_k, \label{AppendBCorr}
\end{eqnarray}%
where $D(\omega_k)$ is the density of states in the leads, and
\begin{equation}
 n _{\operatorname{th}}^{(L)}\left(
\omega\right)\equiv
\frac{1}{1+e^{(\hbar\omega-\mu_{L})/k_{\operatorname{B}}T}},
\end{equation}
is the thermal occupation number of the lead in quasi-equilibrium.
The Markovian approximation requires two assumptions. First assumed is the
``flat band'' condition that the relative change of the
effective density of states around the resonant
$\omega_0$ over a range of the characteristic damping rate
$\gamma_{L}$ is much less than unity, i.e.,
\begin{equation}
\left(\frac{\partial \ln \bar{D}(\omega_k)}{\partial
\omega_k}\right)^{-1}\gg \gamma_{L},
\end{equation}
where $\bar{D}(\omega_k)\equiv\xi^{2}(\omega_k)D({\omega_k})$. Under
this condition, $\bar{D}(\omega_k)$ can be replaced by its value at
$\omega_0$, and the correlation becomes
\begin{eqnarray}
\langle \tilde{\mathcal{L}}^{\dag }\left( t\right)
\tilde{\mathcal{L}}\left( t^{\prime }\right) \rangle
=\bar{D}({\omega_0})\int_{0}^{\mu_{L}}e^{i(\omega _{k}-\omega_0)(
t-t^{\prime }) }d\omega_k. \label{AppendBCorr1}
\end{eqnarray}%
Here, the zero temperature case has been considered for simplicity.
Second, under the large bias condition, the resonant level
$\omega_0$ is far away from the fermi energy and the conduction band
bottom (chosen as the energy origin), i.e.
\begin{equation}
\mu_L-\omega_0, \omega_0\gg\gamma_{L}.
\end{equation}
In this case, the integration over $\omega_k$ is extended to
$\pm\infty$, and finally results in the white-noise correlation

\begin{equation}
\langle \tilde{\mathcal{L}}^{\dag }\left( t\right)
\tilde{\mathcal{L}}\left( t^{\prime }\right)
\rangle=\gamma_L\delta(t-t'),
\end{equation}
where $\gamma _{L}=2\pi \xi ^{2}\left( \omega _{0}\right)D\left(
\omega _{0}\right) $.

Similarly, for the right lead,
\begin{eqnarray}
&& \langle \tilde{\mathcal{R}}\left( t\right)
\tilde{\mathcal{R}}^{\dag }\left( t^{\prime }\right) \rangle 
=\gamma_R\delta(t-t').
\end{eqnarray}
Here, we have use the fact that the thermal occupation number
$n^{(R)} _{\operatorname{th}}\left( \omega _{j}\right)=0$ for the
right lead around the resonant level $\omega_0$. In the same way,
one can show that other noise correlations vanish, i.e.
\begin{equation}
\langle \tilde{\mathcal{L}}\left( t\right) \tilde{\mathcal{L}}^{\dag
}\left( t^{\prime }\right) \rangle= \langle
\tilde{\mathcal{R}}^{\dag }\left( t\right) \tilde{\mathcal{R}}\left(
t^{\prime }\right) \rangle=0.\label{NormalSeq}
\end{equation}

Note that Eq.~(\ref{NormalSeq}) implies that the noise operators
$\tilde{\mathcal{L}}^{\dag }\left( t\right)$ and
$\tilde{\mathcal{R}}\left( t\right)$ play the role of ``annihilation
operators'', since they always give zero correlations when they
stand on the rightmost position. With this observation, the
normal-ordered product of noise operators can be defined by placing
$\tilde{\mathcal{L}}^{\dag }\left( t\right)$ and
$\tilde{\mathcal{R}}\left( t\right)$ to the rightmost position, and
the expectation value of the normal-ordered product vanishes
identically. Thus, the Wick's theorem is generalized to the noise
operators and the current and current noise can be exactly
calculated in the white-noise limit.

\end{appendix}

\begin{acknowledgements}
This work is supported by NSFC No. 10574077, No. 10774085, No.
90203018, No. 10474104, No. 60433050, and No. 10704023, the ``863''
Programme of China No. 2006AA03Z0404, MOST Programme of China No.
2006CB0L0601, NFRPC No. 2006CB921205 and 2005CB724508, Hong Kong RGC
Project 2160322 and Hong Kong RGC Direct Grant 2060346.
\end{acknowledgements}


\begin{thebibliography}{99}

\bibitem{RMPspinQD} R. Hanson, L. P. Kouwenhoven, J. R. Petta, S. Tarucha, and L. M. K. Vandersypen,
Rev. Mod. Phys. \textbf{79}, 1217
(2007).
\bibitem{PRLMarcusTarucha} J. R. Petta, A. C. Johnson, J. M.
Taylor, E. A. Laird, A. Yacoby, M. D. Lukin, C. M. Marcus, M. P.
Hanson, and A. C. Gossard, Science \textbf{309}, 2180 (2005); F. H.
L. Koppens, C. Buizert, K.-J. Tielrooij, I. T. Vink, K. C. Nowack,
T. Meunier, L. P. Kouwenhoven, and L. M. K. Vandersypen, Nature
\textbf{442}, 766 (2006); J. Baugh, Y. Kitamura, K. Ono, and S.
Tarucha, Phys. Rev. Lett. \textbf{99}, 096804 (2007).


\bibitem{PhysReportNoise} Y. M. Blanter and M.~B\"{u}ttiker, Phys. Rep. \textbf{336}, 1
(2000).
\bibitem{PhysTodayShotNoise} C.~W.~J.~Beenakker and C.~Schonenberger, Phys.
Today \textbf{56}, 37 (2003).
\bibitem{PRLMarcusNoise} L. DiCarlo, Y. Zhang, D. T. McClure, D. J. Reilly,
C. M. Marcus, L. N. Pfeiffer, and K. W. West, Phys. Rev. Lett.
\textbf{97}, 036810 (2006); D. T. McClure, L. DiCarlo, Y. Zhang,
H.-A. Engel, C. M. Marcus, M. P. Hanson, and A. C. Gossard, Phys.
Rev. Lett. \textbf{98}, 056801 (2007); Y. Zhang, L. DiCarlo, D. T.
McClure, M. Yamamoto, S. Tarucha, C. M. Marcus, M. P. Hanson, and A.
C. Gossard, Phys. Rev. Lett. \textbf{99}, 036603 (2007).


\bibitem{Landauer} R. Landauer, IBM J. Res. Dev. \textbf{32}, 306 (1988).
\bibitem{bookDatta1}S. Datta, {\it Electronic transport in
mesoscopic systems} (Cambridge University Press, Cambridge, 1995).
\bibitem{bookQK} H. Haug and A.-P Jauho, {\it Quantum kinetics in
transport and optics of semiconductors} (Spinger, Berlin, 1996).
\bibitem{bookDatta2} S. Datta, {\it Quantum transport: Atom to transistor} (Cambridge University
Press, Cambridge, 2005).
\bibitem{Nazarov} T. H. Stoof and Y. V. Nazarov, Phys. Rev. B
\textbf{53}, 1050 (1996); B. L. Hazelzet, M. R. Wegewijs, T. H.
Stoof, and Y. V. Nazarov, Phys. Rev. B \textbf{63}, 165313 (2001).
\bibitem{GurvitzPRBMasterEq} S. A. Gurvitz and Y. S. Prager, Phys. Rev. B \textbf{53}, 15932
(1996).
\bibitem{PRLBrandesA} R. Aguado and T. Brandes, Phys.Rev. Lett.
\textbf{92}, 206601, (2004).
\bibitem{PRLBrandesB} G. Kieszlich, E. Sch\"{o}ll, T. Brandes, F. Hohls, and R. J. Haug, Phys.Rev. Lett. \textbf{99}, 206602
(2007).
\bibitem{Kieblich} G. Kie{\ss}lich, A. Wacker, and E. Sch\'{o}ll, Phys. Rev. B, \textbf{68}, 125320
(2003).
\bibitem{Aghassi}J. Aghassi, A. Thielmann, M. H. Hettler, and G. Sch\"{o}n, Phys.
Rev. B \textbf{73}, 195323 (2006).
\bibitem{LiXinQiCurrent} X.-Q. Li, J. Luo, Y.-G. Yang, P. Cui, and Y.J. Yan, Phys. Rev. B \textbf{71}, 205304
(2005).
\bibitem{LiXinQiNoise} J.Y. Luo, X.-Q. Li, and Y.J. Yan, Phys. Rev. B \textbf{76}, 085325
(2007).
\bibitem{DongBRate} B. Dong, H. L. Cui, and X. L. Lei, Phys. Rev. B \textbf{69}, 035324
(2004); I. Djuric, B. Dong, H. L. Cui, J. Appl. Phys. \textbf{99},
63710 (2006).
\bibitem{DongBLangevin} B. Dong, N. J. M. Horing, and H. L. Cui,
Phys. Rev. B \textbf{72}, 165326 (2005); X. Y. Shen, B. Dong, X. L.
Lei, and N.J.M. Horing, Phys. Rev. B \textbf{76}, 115308 (2007); B.
Dong, X. L. Lei, and N. J. M. Horing, arXiv:0801.0292 (2008).

\bibitem{bookNoise} D. K. C. MacDonald, \textit{Noise and Fluctuations: An Introduction} (John Wiley \& Sons, Inc.,  New York,
1962).

\bibitem{bookGardiner} C. W. Gardiner and P. Zoller, \textit{Quantum Noise} 3rd
ed., (Springer, Berlin, 2004).
\bibitem{bookLouisell} W. H. Louisell, \textit{Quantum Statistical Properties of
Radiation} (John Wiley \& Sons Inc., New York, 1973).
\bibitem{bookHaken} H. Haken, \textit{Light} (North-Holland Pub.
Co., Amsterdam, 1985) Vol. 2 Chapt. 10.
\bibitem{bookScully} M. O. Scully and M. S. Zubairy, \textit{Quantum Optics} (Cambridge University
Press, Cambrdge, 1997).

\bibitem{SuperPoissonBruder} A. Cottet, W. Belzig, and C. Bruder,
Phys. Rev. B \textbf{70}, 115315 (2004); A. Cottet,W. Belzig, and C.
Bruder, Phys. Rev. Lett. \textbf{92}, 206801 (2004).

\bibitem{FSC} W. Belzig, Phys. Rev. B \textbf{71}, 161301(R) (2005).

\bibitem{PRLgreenFunc} Y. Meir, N. S. Wingreen and P. A. Lee, Phys.
Rev. Lett. \textbf{66}, 3048 (1991).

\bibitem{bookAPinteraction} C. Cohen-Tannoudji, J. Dupont-Roc, and
G. Grynberg, \textit{Atom-photon interactions : basic processes and
applications} (John Wiley \& Sons, Inc., New York, 1992)
\bibitem{bookDFWalls} \textit{Quantum Optics}, D. F. Walls and G. J. Milburn,
(Springer-Verlag, New York, 1995).

\bibitem{PapaerGardiner1} C. W. Gardiner and M. J. Collett, Phys. Rev. A \textbf{31}, 3761
(1985).

\bibitem{PapaerGardiner2}  C. W. Gardiner and A. S. Parkins, J. Opt. Soc. Am. B
\textbf{4}, 1683 (1987).

\bibitem{PapaerGardiner3} C. W. Gardiner, cond-mat/0310542.




\end{thebibliography}
\end{document}